\documentclass[journal=jacsat,manuscript=article]{achemso}
\pdfoutput=1 

\usepackage{tikz}
\usepackage{dcolumn}
\usepackage{bm}
\usepackage{upgreek}
\usepackage{array}
\usepackage{booktabs}
\usepackage{microtype}
\usepackage{cmap}
\usepackage{geometry}
\usepackage{siunitx}
\usepackage{hyperref}
\usepackage{amssymb,graphicx}

\usepackage[T1]{fontenc} 
\usepackage[version=3]{mhchem} 

\newcommand{\sqdiamond}[1][fill=black]{\tikz [x=1.2ex,y=1.85ex,line width=.1ex,line join=round, yshift=-0.285ex] \draw  [#1]  (0,.5) -- (.5,1) -- (1,.5) -- (.5,0) -- (0,.5) -- cycle;}

\DeclareSIPostPower\tothefourth{4}
\DeclareSIUnit{\cal}{cal}
\DeclareSIUnit\Molar{\textsc{m}}

\author{John Bridstrup}
\email{john.bridstrup@gmail.com}
\affiliation{Department of Physics, Drexel University, Philadelphia, PA 19104}

\author{John S. Schreck}
\email{jsschreck@gmail.com}
\affiliation{National Center for Atmospheric Research, Boulder, CO 80305}
\affiliation{Department of Chemistry, Drexel University, Philadelphia, PA 19104}

\author{Jesse L. Jorgenson}
\affiliation{The Evernote Corporation, Austin, TX 78705}

\author{Jian-Min Yuan}
\email{yuanjm@drexel.edu}
\affiliation{Department of Physics, Drexel University, Philadelphia, PA 19104}

\title{Stochastic kinetic treatment of protein aggregation and the effects of macromolecular crowding}


\begin{document}

\begin{abstract}
Investigation of protein self-assembly processes is important for the understanding of the growth processes of functional proteins as well as disease-causing amyloids. Inside cells, intrinsic molecular fluctuations are so high that they cast doubt on the validity of the deterministic rate equation approach. Furthermore, the protein environments inside cells are often crowded with other macromolecules, with volume fractions of the crowders as high as $40\%$. We study protein self-aggregation at the cellular level using Gillespie's stochastic algorithm and investigate the effects of macromolecular crowding using models built on scaled-particle and transition-state theories. The stochastic kinetic method can be formulated to provide information on the dominating aggregation mechanisms in a method called reaction frequency (or propensity) analysis.  This method reveals that the change of scaling laws related to the lag time can be directly related to the change in the frequencies of reaction mechanisms. Further examination of the time evolution of the fibril mass and length quantities unveils that maximal fluctuations occur in the periods of rapid fibril growth and the fluctuations of both quantities can be sensitive functions of rate constants. The presence of crowders often amplifies the roles of primary and secondary nucleation and causes shifting in the relative importance of elongation, shrinking, fragmentation and coagulation of linear aggregates.  Comparison of the results of stochastic simulations with those of rate equations gives us information on the convergence relation between them and how the roles of reaction mechanisms change as the system volume is varied.   

\end{abstract}

\section{TOC Graphic}
\begin{figure}[h]
\centering
\includegraphics[width=3.25 in]{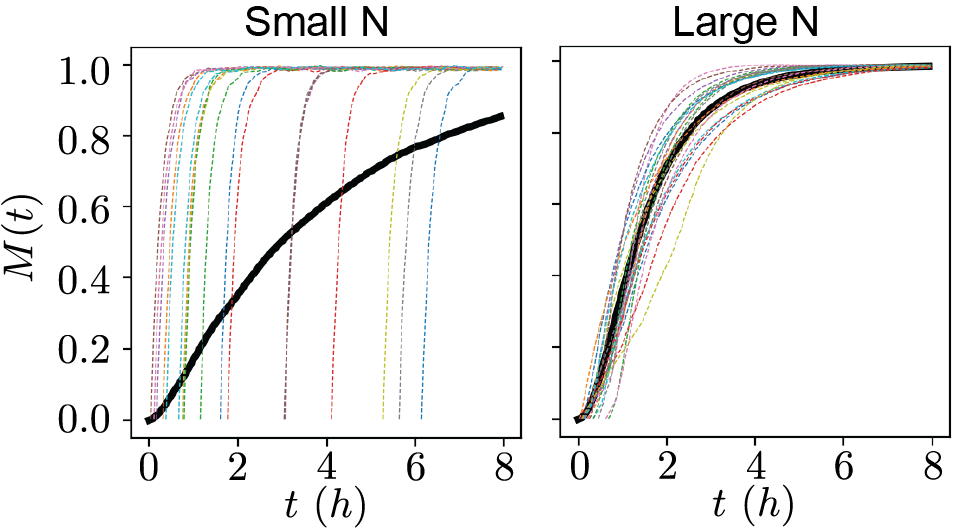}
\end{figure}

\section{Introduction}

Protein self-assembly is an important process for the formation of natural polymers like actin filaments and microtubules\cite{actin,microtubule}, but also for the formation of amyloids\cite{Knowles-PhysToday,yuan17}, culprits for many neurodegenerative diseases including Alzheimer's and Parkinson's diseases. Decades of studies of self-assembly of proteins reveal that the molecular reactions involved can be complicated, for example, in the processes of amyloid formaton.\cite{knowles-09,michaels14,hong17,urbanc17}  

Among the many theoretical methods to investigate the protein aggregation problem, the rate equation approaches based on mass-action laws\cite{frieden07,murphy07,gillam13,knowles-09,Schreck13,michaels14,hong17} often provide direct fits to experimental data and interpretation of the aggregation processes. Processes considered in these studies often include primary nucleation, monomer addition and subtraction, fibril fragmentation, merging of oligomers, heterogeneous (or surface-catalyzed) nucleation, etc.  The reaction rates associated with the reaction steps considered can be independent or dependent on the oligomer/fibril size.\cite{Schreck13,michaels14,hong17}

Almost all of our current knowledge comes from the studies of systems $\textit{in vitro}$, and little has been done in understanding processes $\textit{in vivo}$. In the latter case, the volume of the compartment or inside confining boundaries is usually small and the numbers of copies of certain protein species can be low, resulting in large number fluctuations \cite{ferrone80,hofrichter,knowles11,wittrup06,michaels16,Szavits14,tiwari16}. Furthermore, experiments carried out on protein aggregation often starts out with monomers of amyloid peptides or proteins. This may be what happens in the brains also, starting with monomers.  Above the critical concentration, monomers then aggregate into dimers, trimers, tetramers, \dots, by steps. The system necessarily goes through the stages in which the numbers of dimers, trimers, \dots, oligomers are very small. On the other hand, deterministic rate equations based on mass-action laws are often used to simulate the growth of the oligomer species. Rigorously speaking, rate equations, that is, concentrations, are well-defined only when large numbers of relevant molecules exist, for the fluctuations are inversely proportional to the square root of the number of molecules. Therefore, application of the rate equations to the cases involving small numbers of oligomers cannot fully be justified. For these cases, one should use stochastic dynamics methods, such as the Gillespie algorithm.\cite{gillespie76, gillespie77, gillespie92, gillespie07, gibson00, mcquarrie67} From another point of view, it would be important to evaluate the accuracy of the rate equation results, especially in the early stage of aggregation, or at any times when any numbers of the chemical species involved are small, by comparing them to those obtained using a stochastic kinetic method. 

Gillespie's method of carrying out a stochastic chemical kinetics study is to solve the master equations defined on the probabilities, $\bold{P}(N_1, N_2, N_3,\dots, N_k,\dots, N_M)$, where $N_k$ is the number of $k$-mer, $M_k$, denoting an oligomer containing $k$ monomers. We can estimate the dimension, $D$, of the vector space of $\bold{P}$. Consider an experiment for which only $M$ monomers exist initially at $t$ = 0, that is, $N_k(0) = 0$, for $k \ge 2$. Since $N_1$ is bound above by $M$, $N_2$ by $M/2$, $N_k$ by $M/k$, and $N_M$ by 1, the dimension of the vector $\bold{P}$ has an upper bound given by $M^M/(M!)$. Thus for a large $M$, $D$ is bounded above by $e^M$, using the Sterling approximation. This implies that the dimension of the state space grows very fast as $M$ increases. We cannot realistically integrate the master equations numerically for an $M$ larger than a few hundreds. In reality, for amyloid fibrils, $M$ can go as high as several thousands. Thus currently the standard stochastic kinetic method is restricted to investigating the early stage of the protein self-assembly processes or, alternatively, to investigating systems of small volume, in which the total number of proteins in the system is relatively small.

The fact that the stochastic kinetic methods can be used to study chemical reactions in small volumes was pointed out early on by Gillespie. \cite{gillespie76, gillespie77, gillespie92, gillespie07} Applications of the stochastic kinetic methods to study protein self-assembly processes in small volumes have been carried out recently by several groups. Szavits-Nossan, et al. have derived an analytic expression for the lag-time distribution based on a simple stochastic model, including in it primary nucleation, monomer addition, and fragmentation as possible reaction mechanisms.\cite{Szavits14} Tiwari and van der Schoot have carried out an extensive investigation of nucleated reversible protein self-aggregation using the kinetic Monte Carlo method.\cite{tiwari16} They focused on the stochastic contribution to the lag time before polymerization sets in and found that in the leading order the lag time is inversely proportional to system volume for all nine different reaction pathways considered. Michaels, et al.\ on the other hand, have carried out a study of protein filament formation under spatial confinement using stochastic calculus, focusing on statistical properties of stochastic aggregation curves 
and the distribution of reaction lag time.\cite{michaels18}

At the cellular level, besides the reaction being contained in a small volume, the environments of proteins are crowded with other biomolecules, such as DNA, lipids, other proteins, etc. The fraction of volume occupied by these ``crowders'' can be as high as $30$ - $40\%$, which can affect the reaction rates of proteins as well as other biopolymers in the cell in significant ways.\cite{bridstrup16,hall02,hall04,zhou08,schreck17,schreck20,hong2020understanding}

In this article, we extend the earlier stochastic works on protein self-assembly by investigating beyond the early stage of aggregation to include the polymerization phase and present a general study of fluctuations for systems in small volumes (i.e., low total number of proteins). Furthermore, we examine the role of macromolecular crowding in changing the microscopic behavior of an aggregating system, and present a new method for extracting the dominant aggregation mechanisms that is more explicit than scaling law analysis.\cite{meisl16-prot,michaels16} We also compare the stochastic and rate-equation approaches for a model case to gain insight on when and why the methods begin to diverge. 

The organization of the present article is as follows: In Section II we describe the kinetic models used in the study. In Section III we introduce the master equation, present Gillespie's stochastic simulation algorithm (SSA), discuss how it connects to a rate equation treatment, and finally introduce our reaction frequency method of analysis. In Section IV we give a brief overview of how the effects of macromolecular crowding are included in the models through scaled-particle and transition-state theories. Our results are presented in Section V. Finally, we conclude with discussion and comments in Section VI.

\section{Kinetic Models Studied}

\subsection{The Oosawa Model}

\begin{figure*}[ht]
    \centering
    \includegraphics[width=7 in]{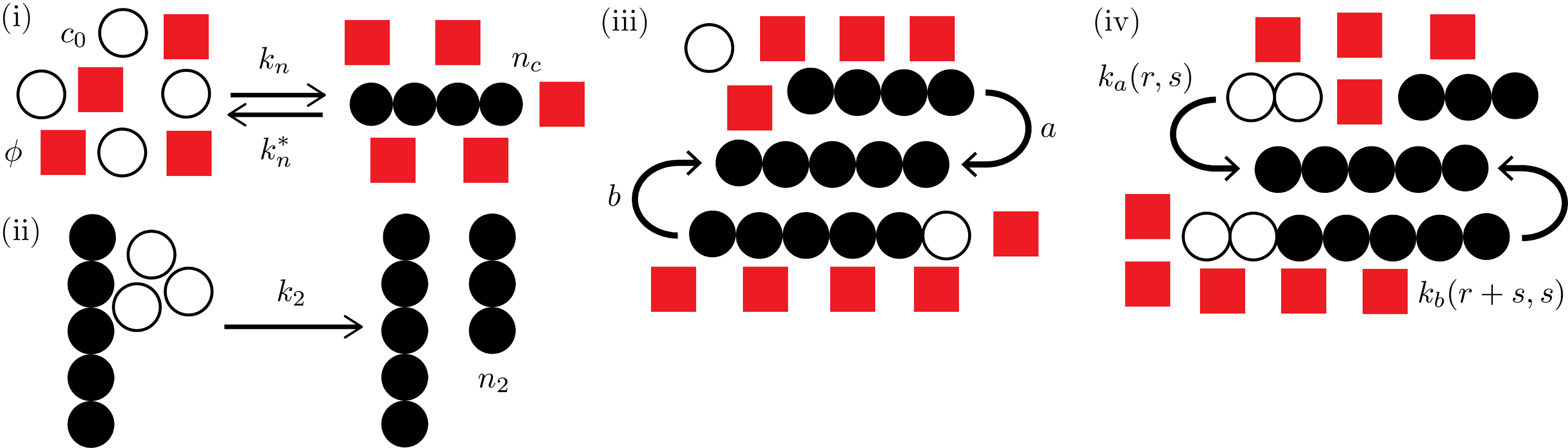}
    \caption{Mechanisms of protein aggregation included in this study. (i) Primary nucleation, (ii) one-step secondary nucleation, (iii) monomer addition and subtraction, (iv) coagulation and fragmentation.}
    \label{fig:basic}
\end{figure*}

The Oosawa model\cite{oosawa} is a simple and widely studied\cite{bridstrup16, Wattis, BD, hall04} model of protein filament formation. Protein aggregates of size $n_c$, the \textit{critical nucleus} size, form initially by primary nucleation, as is illustrated schematically in Fig.~\ref{fig:basic}(i), and grow in one possible way by simple monomer addition and subtraction, shown in Fig.~\ref{fig:basic}(iii). It is assumed based on the classical nucleation theory  that the concentrations of aggregates smaller than $n_c$ are zero. The model is described by the kinetic equations
\begin{align}
\label{eq:oosawa}
    \frac{dc_{n_c}}{dt} &= k_nc_1^{n_c} - (ac_1 + k_n^*)c_{n_c} + bc_{n_c+1}\nonumber\\
    \frac{dc_{r>n_c}}{dt} &= ac_1(c_{r-1} - c_r) + b(c_{r+1} - c_r)\\
    \frac{dc_1}{dt} &= n_c(k_n^{*}c_{n_c} - k_nc_1^{n_c})-\sum_{r=n_c+1}^{\infty}r\frac{dc_r}{dt}\nonumber
\end{align}
where $a$, $b$ and $k_n$ are, respectively, the monomer addition, monomer subtraction and primary nucleation rate constants, $k_n^*$ is the rate constant for a nucleus dissolving and $c_i$ is the concentration of an oligomer containing i monomers, denoted by  $M_i$. There are several systems, such as the growth of actin,\cite{bridstrup16, schreck17, oosawa} for which this model describes experimental data quite well.

\subsection{Generalized Smoluchowski Model}

The generalized Smoluchowski model\cite{Schreck13, schreck17, schreck20} extends the Oosawa model by allowing for protein aggregates to break and merge with each other. This is a more general and thus more widely applicable model of protein aggregation behavior. To account for breaking and merging in the kinetic equations, Eq.~(\ref{eq:oosawa}) is modified in the following way
\begin{align}
    \label{eq:smol}
    \frac{dc_r}{dt} = ... &+ k_a\Bigg[\frac{1}{2}\sum_{s=n_c}^{r-n_c}c_sc_{r-s} - \sum_{s=n_c}^{\infty}[1 + \delta_{rs}]c_rc_s\Bigg]\nonumber\\
    &+k_b\Bigg[\sum_{s=2}^{\infty}[1 + \delta_{rs}]c_{r+s} - (r - 3)c_r\Bigg],\\
    \frac{dc_1}{dt} =  n_c&(k_n^{*}c_{n_c} - k_nc_1^{n_c})-ac_1\sum_{r=n_c}^{\infty}c_r+b\sum_{r=n_c+1}^{\infty}c_r\nonumber\\
    &+k_b\Bigg[\sum_{s=n_c}^{\infty}\sum_{r=2}^{n_c-1}(1+\delta_{rs})rc_{r+s}\Bigg]\nonumber
\end{align}
where $k_a$ and $k_b$ are the coagulation and fragmentation (Fig. \ref{fig:basic}(iv)) rate constants, respectively, $\delta_{rs}$ is the Kronecker delta function, and the factors of $\frac{1}{2}$ and $r-3$ take into account double counting and aggregates having multiple points at which they can break. The last term on the RHS of the monomer concentration equation is due to the assumption that any polymers smaller than $n_c$ which form via fragmentation immediately dissolve into monomers. 

\subsection{Secondary Nucleation}

In some cases, an additional mechanism is needed to accurately describe aggregation: secondary or heterogeneous nucleation. Secondary nucleation is the process of existing polymers catalyzing the formation of new aggregates on their surface and has been shown\cite{Ferrone02,Ferrone85, ferrone80} to play an important role in many aggregating systems. We investigate a simple model of secondary nucleation, where the process is modelled as a one-step process, shown in Fig. \ref{fig:basic}(ii). In reality, secondary nucleation can be more generally represented as a two-step process, which reduces to a one-step process in the low monomer concentration limit. \cite{meisl14,schreck20} But for the present purposes, we will focus on the one-step secondary nucleation process.

For one-step secondary nucleation, the rate equations in \ref{eq:oosawa} and \ref{eq:smol} are modifed by
\begin{equation}
\label{eq:one-step}
    \frac{dc_{n_2}}{dt} = ... + k_2c_1^{n_2}M_{free},
\end{equation}
where $n_2$ is the secondary nucleus size, and $M_{free}$ is the total mass of  the polymers which contain $n_2$ or more monomers. 


\section{Stochastic Chemical Kinetics}

In classical chemical kinetics, it is assumed that the concentrations of chemical species vary continuously over time, the so-called mean-field approach.\cite{michaels14} While this is often appropriate for bulk systems, where volume can be assumed to be macroscopic and populations are roughly of the order $N_A$, Avogadro's number, it does not take into account that species populations are discrete and change in integer numbers. Further, chemically reacting systems display random (stochastic) behavior due to their contact with a reservoir. Stochastic fluctuations can become increasingly relevant as system size or overall concentration decrease. Thus, the assumption that populations can be represented as continuous concentrations becomes less valid as the number of molecules becomes smaller. Stochastic chemical kinetics aims to describe how a well-stirred reacting system evolves in time, taking into account fluctuations and the discreteness of populations. The primary goal of stochastic kinetics is to solve the chemical master equation (CME)\cite{gillespie92}
\begin{align}
\label{eq:cme}
    \frac{\partial P(\textbf{x}, t|\textbf{x}_0, t_0)}{\partial t} = &\sum_{j=1}^{S} [a_j(\textbf{x}-\textbf{v}_j)P(\textbf{x}-\textbf{v}_j, t | \textbf{x}_0, t_0)\nonumber\\
    &-a_j(\textbf{x})P(\textbf{x}, t | \textbf{x}_0, t_0)],
\end{align}
where $\textbf{x}$ is the state of the system, $\textbf{v}_j$ is change in system state due to a single reaction $R_j$, $P(\textbf{x}, t|\textbf{x}_0, t_0)$ is the probability of the system being in state $\textbf{x}$ at time $t$ given an initial state $\textbf{x}_0$ at time $t_0$, and $a_j(\textbf{x})$ is the propensity function, or transition rate, of a given reaction $R_j$ defined by
\begin{align}
\label{eq:propensity}
    &a_j(\textbf{x})dt \triangleq \mbox{ the probability, given }\textbf{x}\mbox{,} \mbox{ that one }\\&R_j\mbox{ reaction occurs in the time interval }dt.\nonumber
\end{align}
We will show later that the propensity function can be directly related to the reaction rates from classical chemical kinetics.

In principle, the probability distribution, $\textbf{P}(\textbf{x}, t|\textbf{x}_0, t_0)$ is entirely described by eq. \ref{eq:cme}. In practice, however, analytical solutions are often impossible due to the CME being, in general, a very large system of coupled ODEs. Other complications with a direct analysis of the CME are described by Gillespie\cite{gillespie92,gillespie07}. Thus, it is necessary to use computational methods to solve for the evolution of the probability distribution function. Several methods of simulating exact and approximate solutions to the CME have been proposed.\cite{gillespie07, padgett16} In our study we use the Gillespie stochastic simulation algorithm (SSA), which allows for a highly detailed, albeit somewhat computationally expensive, look at how a stochastic system evolves over time. 

\subsection{Gillespie Simulation Algorithm}

The Gillespie stochastic simulation algorithm is a method for generating statistically accurate reaction pathways of stochastic equations, and thus statistically correct solutions to the CME (eq. \ref{eq:cme}). The algorithm proceeds as follows:

\begin{enumerate}
    \item Set the species populations to their initial values and $t=0$.
    \item Calculate the transition rate, $a_i(\textbf{x})$, for each of the $S$ possible reactions.
    \item Set the total transition rate $Q(\textbf{x}) = \sum_{j=1}^{S}a_j(\textbf{x})$.
    \item Generate two uniform random numbers, $u_1$ and $u_2$.
    \item Set $\Delta t = \frac{1}{Q(\textbf{x})}ln(\frac{1}{u_1})$.
    \item Find $\mu \in [1,..,S]$ such that \\
    $\sum_{j=1}^{\mu -1}a_j(\textbf{x}) < u_2Q(\textbf{x}) \leq \sum_{j=1}^{\mu}a_j(\textbf{x})$.
    \item Set $t = t + \Delta t$ and update species populations based on reaction $\mu$.
    \item Return to step 2 and repeat until an end condition is met.
\end{enumerate}

This process generates a single reaction pathway and may be repeated and averaged to compare with bulk behavior and experimental results. Modifications to this method exist to decrease computation time, such as the $\textit{next-reaction}$\cite{gibson00} and $\textit{tau-leaping}$\cite{padgett16} methods, but where they improve efficiency they sacrifice in accuracy.

\subsection{Relation to Chemical Kinetics}

In classical chemical kinetics, differential rate equations are solved to study the bulk behavior of a continuous system. In order to compare with these studies, as well as with experimental studies, it is necessary to relate bulk rates, and rate constants, with the stochastic propensity functions and the equivalent stochastic rate constants. We give two examples of how this is done. For a coagulation process $M_r + M_s \overset{k_a}{\rightarrow} M_{r+s}$, we have a rate equation of the form
\begin{equation}
    \frac{dc_{r+s}}{dt} = k_ac_rc_s.
\end{equation}
Making use of the relationship between species population and species concentration, $c_i = \frac{N_i}{N_AV}$, where $N_i$ is the species population and $V$ is system volume, we find the relation
\begin{equation}
    \frac{dN_{r+s}}{dt} = \frac{k_a}{N_AV}N_rN_s,
\end{equation}
which is the rate at which an $r$-mer and an $s$-mer transition into an $(r+s)$-mer. The RHS is the sum of the propensity functions from eq. \ref{eq:propensity} for all possible $M_r + M_s \rightarrow M_{r+s}$ coagulation reactions. Finally, the stochastic rate constant can be written
\begin{equation}
    k_a' = \frac{k_a}{N_AV}.
\end{equation}
For protein aggregation, we can again make use of the definition of concentration for the total number, $N$, and concentration, $c_0$, of monomers to obtain
\begin{equation}
    k_a' \equiv \frac{c_0}{N}k_a.
\end{equation}
For a primary nucleation process, $n_c M_1 \overset{k_n}{\rightarrow} M_{n_c}$, we have a rate equation
\begin{equation}
    \frac{dc_{n_c}}{dt} = k_nc_1^{n_c}.
\end{equation}
Following the same process, the stochastic nucleation rate constant is given by
\begin{align}
    k_n' &= \frac{k_n}{(N_AV)^{n_c-1}}\nonumber\\ \\
    &\equiv \bigg(\frac{c_0}{N}\bigg)^{n_c-1}k_n.\nonumber
\end{align}
The other stochastic rate constants are found similarly. It is worth noting that all stochastic rate constants have units of frequency ($s^{-1}$) and thus the stochastic rate constants involved in shrinking or breaking processes ($b'$, $k_b'$, $\Bar{k_2}'$, etc.) are identical in value to their bulk counterparts.

\subsection{Reaction Frequency Method}

In numerical simulations using Gillespie's stochastic algorithm, we can obtain information on what reactions are occurring in certain intervals of time. In particular, we investigate the frequencies  or propensities of particular reaction types as they evolve over time. This approach gives unique insight into the behavior of a system and offers direct confirmation of which mechanisms of reaction are dominating at various phases of the aggregation process. For instance, a common scenario, meaning for a set of parameters showing nontrivial dynamic behaviors, is that the primary nucleation dominates at the very beginning, then monomer addition and oligomer coagulation become important, balanced by monomer subtraction and fragmentation. These are followed by secondary nucleation, which becomes active, before monomers are depleted.  At longer time scale, oligomer coagulation and fragmentation persist as the system approaches an equilibrium or steady state.  For some aggregation reactions, such as that involving actin, secondary nucleation never plays a significant role, so it can be neglected. But, for other reactions, especially under the influence of molecular crowders, secondary nucleation dominates, until monomers are depleted. In the results section, we normalize the reaction frequencies by the total number of reactions occurring at that time ($f_{rel}$). Doing this allows us to investigate the relative importance of any particular reaction as it evolves over time.

\section{Macromolecular Crowding}

In the presence of molecular crowders, the rate constants of reaction steps may be affected and the degree of influence varies widely among the different types of reaction steps involved. The effects of crowders on the rate constants has been worked out in our previous study\cite{bridstrup16,schreck20} using the transition-state theory\cite{eyring} (TST) and the scaled-particle theory\cite{cotter74,hall02, schreck17} (SPT). In this section, we outline the relevant formulas that we have used in the present simulations. We start with the forward coagulation reaction of the reversible reaction
\begin{equation}
    \ce{M_r + M_s <=>[k_a][k_b] M_{r+s}}.
\end{equation}
In TST one assumes that quasi-equilibrium is established between the reactants and the transition state which allows us to express the rate constant in terms of the free energy difference between the reactants and the transition state.  Further, expressing the chemical potential of a chemical species in terms of the product of its activity coefficient, $\gamma$, and concentration, we can describe the forward rate constant using the following relationship
\begin{align}
    k_a &= \frac{\gamma_{r}\gamma_{s}}{\gamma_{r+s}^{\ddagger}}k_a^0\nonumber\\
    &\approx \frac{\gamma_{r}\gamma_{s}}{\gamma_{r+s}}k_a^0\\
    &= \frac{\gamma_1}{\alpha}k_a^0,\nonumber
\end{align}
where $\gamma_r$ is the activity coefficient for an $r$-mer, $\ddagger$ refers to the transition state, and the superscript $k_a^0$ refers to the associate rate constant in the case that all relevant activity coefficients are unity, including the case where the crowders are absent in the solution. Thus  we call it the \textit{crowderless} rate constant below. We have also made use of the important result\cite{cotter74,bridstrup16,hall02, Minton-14} that the activity coefficient of an $r$-mer can be related to that of a monomer by $\gamma_r=\gamma_1\alpha^{r-1}$, where the parameter $\alpha$ is derived from the change in activity as a monomer is added to a spherocylinder aggregate. Additionally, the approximation has been made that the transition state has roughly the same structure as the aggregated polymer, thus $\gamma_{r+s}^{\ddagger} \approx \gamma_{r+s}$. This approximation, when applied to the reverse reaction, gives the result
\begin{equation}
    k_b=k_b^0.
\end{equation}
In other words, breaking reactions are unaffected by crowders in this model. This approach can be applied to the other mechanisms in our model to obtain
\begin{align}
    \label{monomer_add_rate}
    a &= \frac{\gamma_1}{\alpha}a^0\nonumber\\
    k_n &= \bigg(\frac{\gamma_1}{\alpha}\bigg)^{n_c-1}k_n^0\\
    k_2 &= \gamma_1^{n_2}\Gamma k_2^0 \nonumber\\
    b & = b^0,\nonumber
\end{align}
where $\Gamma$ is a factor related to the change in shape of an aggregate as monomers attach to the surface in a secondary nucleation process. For our study, we assume $\Gamma \equiv 1$\cite{Ferrone85} in a one-step secondary nucleation model. The activity coefficients, as well as $\alpha$, may be calculated using SPT by treating crowders and monomers as hard spheres and aggregates as hard sphero-cylinders\cite{bridstrup16,Minton-14,schreck20}. Expressions for the effects of crowders on the activity coefficients in this case have been derived by Cotter\cite{cotter74}. Further details on the relevant calculations have been presented by Bridstrup and Yuan\cite{bridstrup16}, and Minton\cite{Minton-14}, and a closer look at how crowders contribute to both the one-step and two-step secondary nucleation has been presented by Schreck et al.\cite{schreck20}

\section{Results}

\subsection{Scaling Laws}

\begin{figure}
    \centering
    \includegraphics[width=3.5 in]{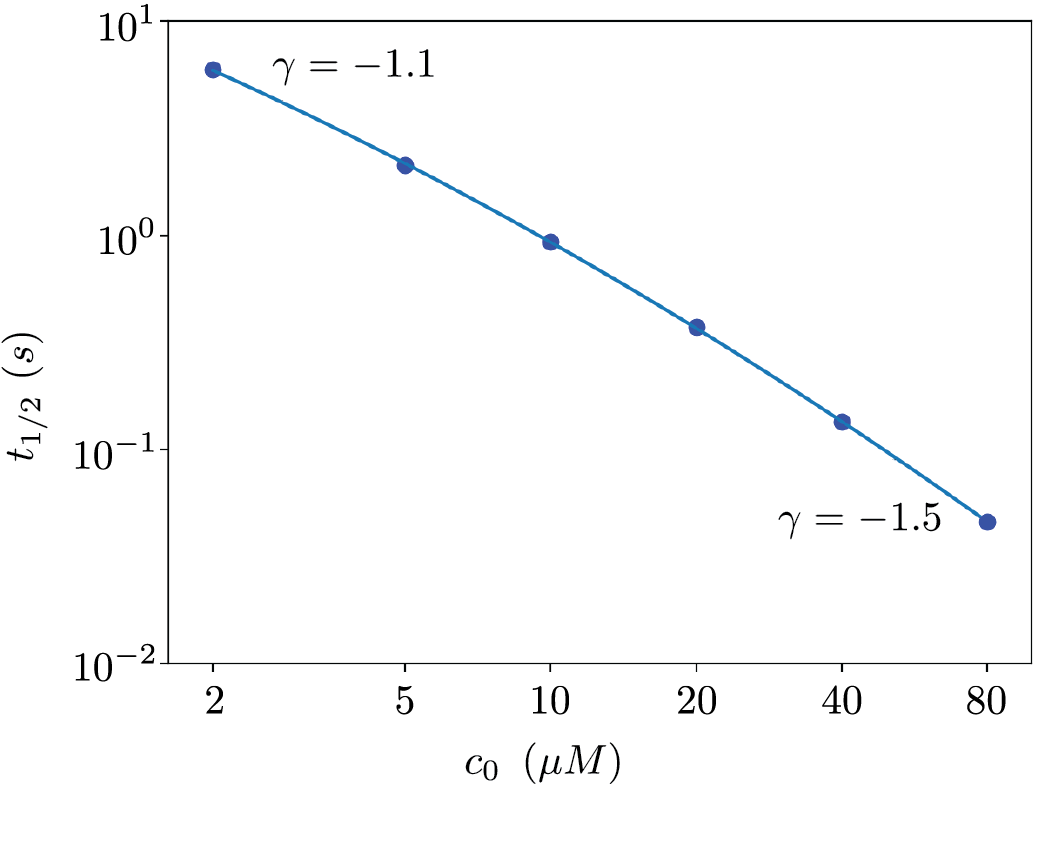} 
    \caption{Log-log plot of $t_{1/2}$ vs $c_0$. At low $c_0$, nucleation-elongation dominates based on the scaling factor $\gamma=-1.1$. At higher concentrations, the scaling exponent reaches $\gamma=-1.5$ corresponding to secondary nucleation being the dominant mechanism. These interpretations are based on the scaling laws of Meisl et al.\cite{meisl16-prot}}
    \label{fig:logc_logt}
\end{figure}

\begin{figure}
    \centering
    \includegraphics[width = 3.5 in]{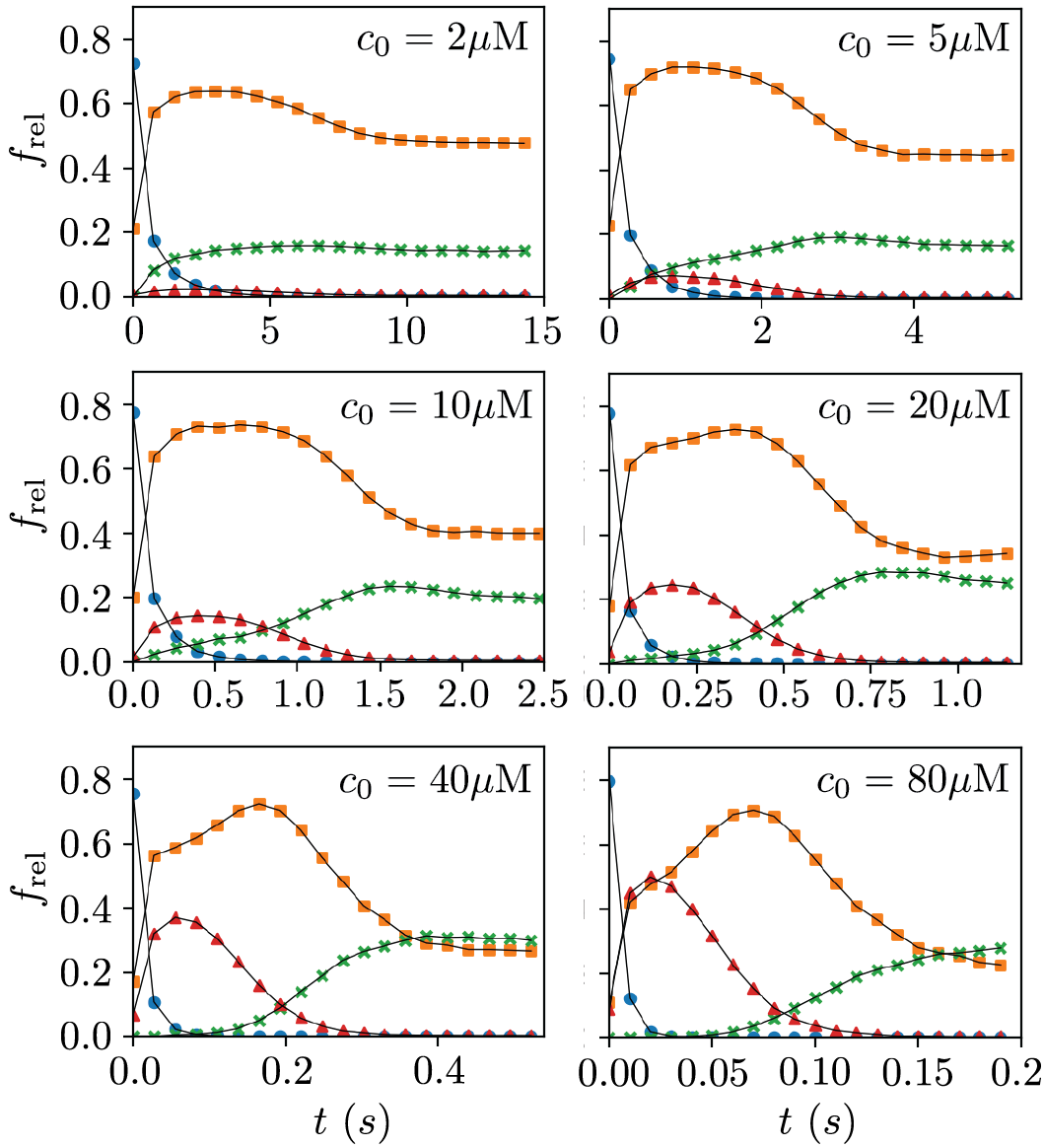}
    \caption{Relative reaction frequencies vs time. Blue ($\bullet$) corresponds to primary nucleation, orange ($\blacksquare$) to monomer addition, red ($\blacktriangle$) to secondary nucleation and green (\textbf{x}) to fragmentation. As $c_0$ is increased, it is clear that secondary nucleation goes from hardly participating in the reaction to completely dominating. Relative frequencies are calculated by dividing the individual reaction rates with the total reaction rate, which includes monomer subtraction and coagulation (not shown in figure).}
    \label{fig:reac_scaling}
\end{figure}

A powerful tool for extracting the dominant mechanisms of aggregation is by looking at how the half-time of aggregation mass ($t_{1/2}$) scales with increasing initial concentration of monomers.\cite{meisl16-prot, meisl17} We test this method using stochastic simulations by directly examining which reactions are dominating during the growth phase. In a scaling law analysis, the slope of a log-log plot of $t_{1/2}$ vs $c_0$ gives the scaling exponent, $\gamma$, which can be related to specific mechanisms of growth. The reader may refer to Fig. 6 from Meisl et al.\cite{meisl16-prot} for a list of some of these scaling relationships. Fig. \ref{fig:logc_logt} shows a scaling relationship plot for the generalized Smoluchowski model with one-step secondary nucleation. The negative curvature and values of the scaling exponent indicate competition between nucleation-elongation (small $c_0$) and secondary nucleation (large $c_0$), based on Meisl et al. Fig. \ref{fig:reac_scaling} shows the relative reaction frequency of each mechanism of growth as they evolve over time. For low $c_0$, it is clear that monomer addition following an initial phase of primary nucleation is the dominant mechanism of growth. As $c_0$ is increased, the relative frequency of both monomer addition and secondary nucleation increase (competition) before eventually secondary nucleation begins to suppress even monomer addition during the growth phase. This comparison both supports the validity of the scaling laws and justifies further use of this approach, as it is clear that much can be gained from this level of detail.

\subsection{Fluctuations and System Volume}

\subsubsection{Half-time Fluctuations}

\begin{figure}
    \centering
    \includegraphics[width = 6 in]{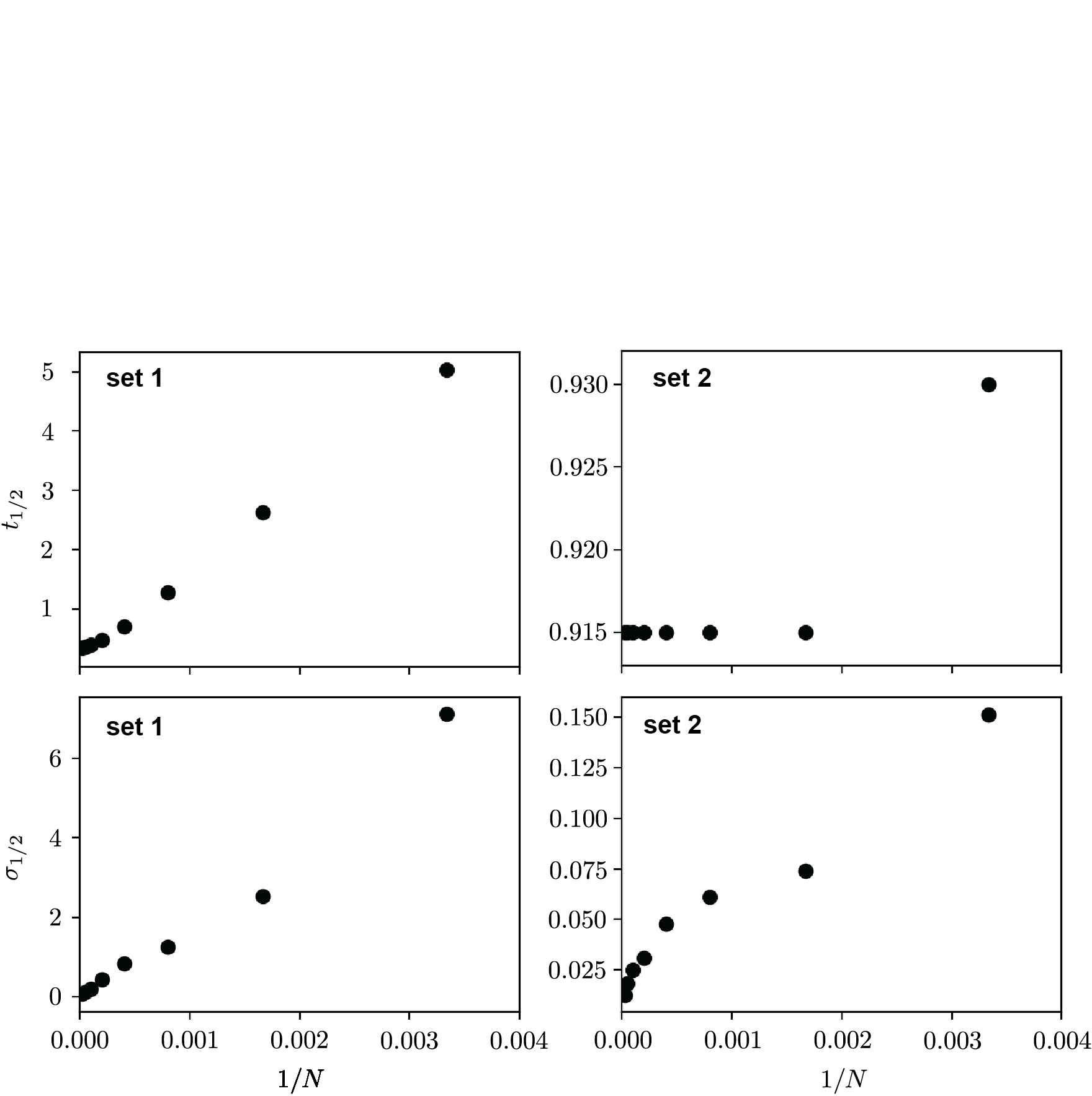}
    \caption{$t_{1/2}$ and $\sigma_{t_{1/2}}$ as functions of $1/N$ for two different sets of rate constants. For \textbf{set 1}, the half-time increases linearly as expected. For \textbf{set 2}, however, there is no increase in half-time until a threshold is reached. Parameters used are, for \textbf{set 1}: $a = \SI{5e3}{\micro \Molar^{-1}s^{-1}}$, $b=k_b=\SI{1e-3}{s^{-1}}$, $k_a=\SI{100}{\micro\Molar^{-1}s^{-1}}$, $k_n=\SI{1e-4}{\micro\Molar^{-1}s^{-1}}$, $n_c=2$, and $c_0=\SI{10}{\micro\Molar}$; and for \textbf{set 2}: $a=\SI{2}{\micro\Molar^{-1}s^{-1}}$, $b=k_b=\SI{1}{s^{-1}}$, $k_a=\SI{0.2}{\micro\Molar^{-1}s^{-1}}$, $k_n=\SI{5e-3}{\micro\Molar^{-1}s^{-1}}$, $n_c=2$, and $c_0=\SI{5}{\micro\Molar}$.}
    \label{fig:th_vs_N-1}
\end{figure}

Tiwari and van der Schoot\cite{tiwari16} showed that nucleation time increases linearly with $1/V$, the inverse of system volume. We investigate the effects of decreasing the volume, or total number of monomers, $N$, at fixed concentration, on the the halftime, $t_{1/2}$, and fluctuations of the halftime, $\sigma_{1/2}$ (the standard deviation of $t_{1/2}$), of polymer mass for two different sets of parameters, which we call \textbf{set 1} and \textbf{set 2} (defined in Fig. \ref{fig:th_vs_N-1}). We define halftime as the time it takes for the polymer mass to reach half its equilibrium value. Fig. \ref{fig:th_vs_N-1} shows $t_{1/2}$ vs $1/N$ for both sets at fixed $c_0$. For \textbf{set 1}, the linear trend is clear at all values of $N$. For \textbf{set 2}, however, there is no increase in half-time until a threshold value is reached. This may imply that bulk behavior is recovered at different values of $N$ depending on the rate constants.

Perhaps unsurprisingly, the relative deviations increase as $N$ decreases. More interestingly, the manner in which these fluctuations increases depends on the rate constants themselves. For \textbf{set 1}, where $t_{1/2}$ increases linearly with $1/N$, the standard deviation ($\sigma_{1/2}$) also increases roughly linearly, whereas for \textbf{set 2}, $\sigma_{1/2}$ appears to increase more like the square root of $1/N$ before abruptly increasing at the same threshold.


\subsubsection{Moment Fluctuations}

\begin{figure}
    \centering
    \includegraphics[width = 3.5 in]{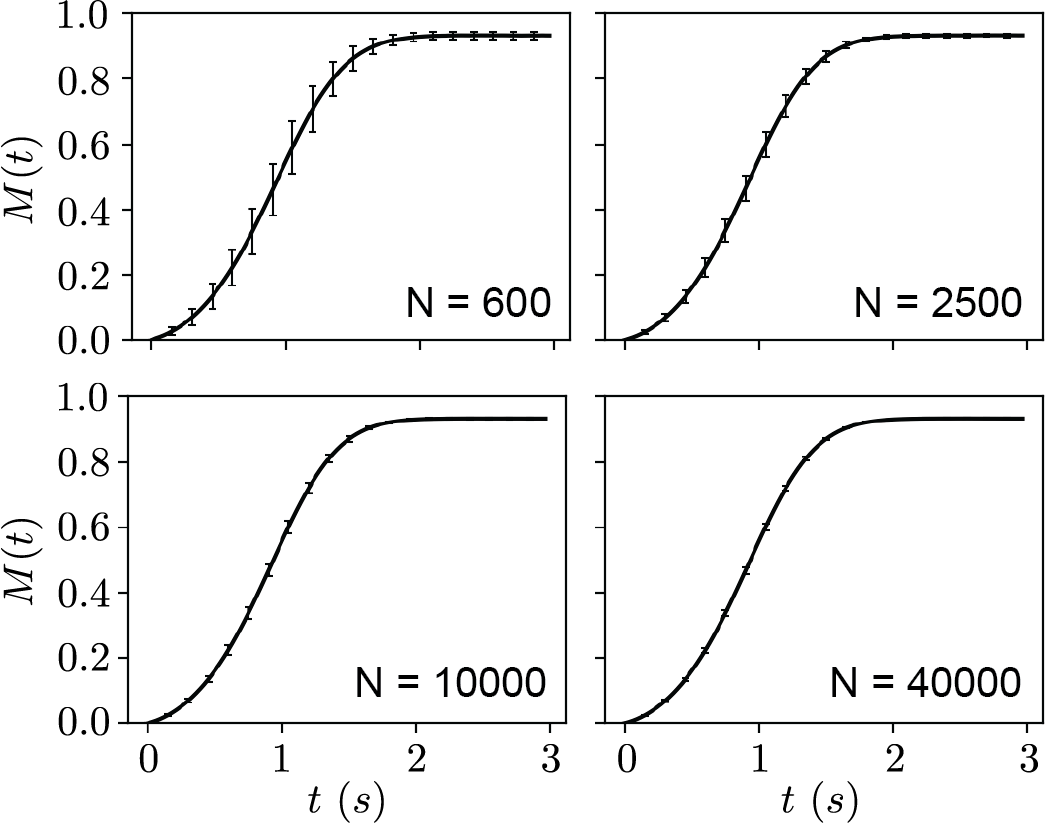}
    \caption{Average mass as a function of time for various values of $N$ using \textbf{set 2} parameters. It is clear that the relative fluctuations decrease significantly as $N$ is increased, proportional to $1/\sqrt{N}$, and also that the time scale for each value of $N$ is the same. Error bars correspond to $\pm1$ standard deviation.}
    \label{fig:mass_dev}
\end{figure}

\begin{figure}
    \centering
    \includegraphics[width = 3.5 in]{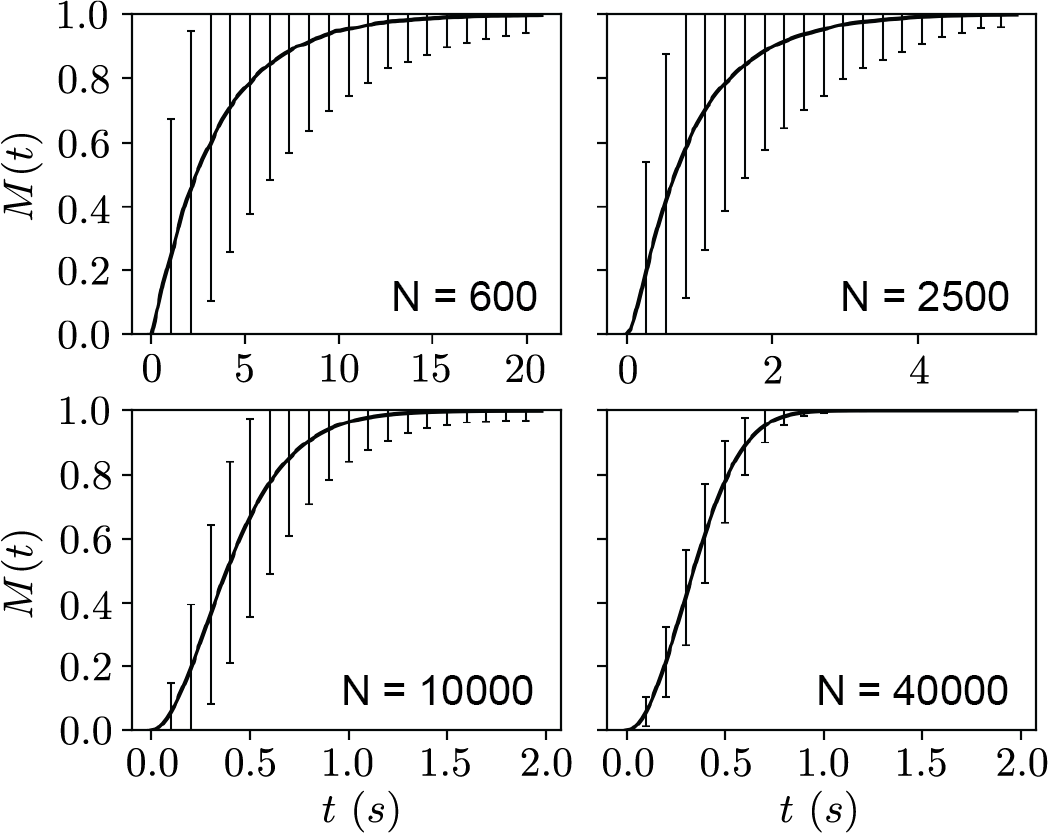}
    \caption{For \textbf{set 1} parameters, $M(t)$ behaves similarly on average for large $N$ but has significantly larger fluctuations. For smaller N, $M(t)$ takes over 10 times as long to reach equilibrium and fluctuations can be up to $100\%$ of the average value for small $N$. Error bars correspond to $\pm1$ standard deviation.}
    \label{fig:bigfluc}
\end{figure}

\begin{figure}
    \centering
    \includegraphics[width = 5 in]{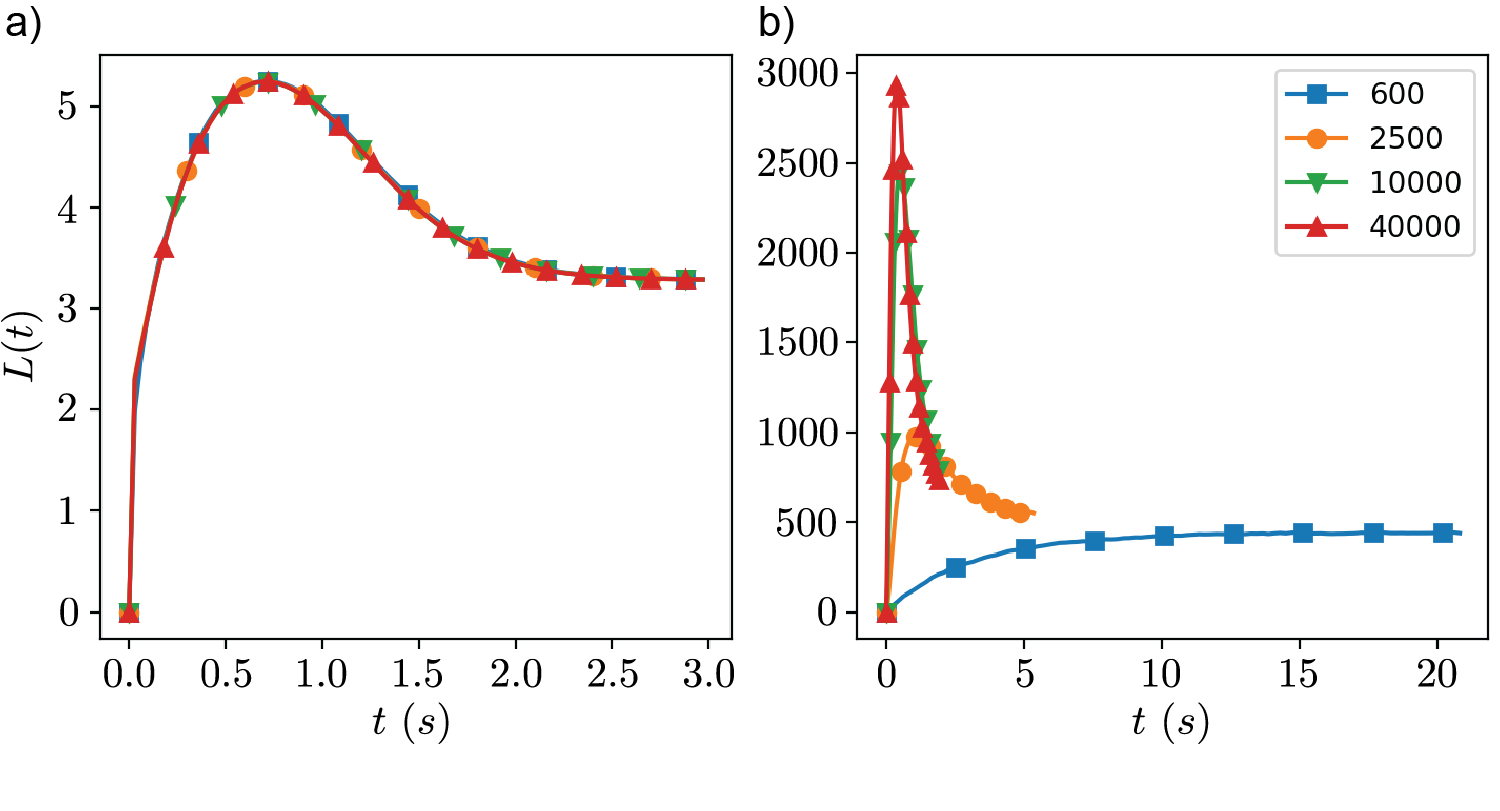}
    \caption{A comparison of $L(t)$ for the two sets of rate constants. For \textbf{set 2} shown in (a), $L(t)$ is almost exactly the same even for very small $N$. For \textbf{set 1} shown in (b), however, $L(t)$ not only peaks lower but it's overall behavior changes significantly as $N$ is decreased.}
    \label{fig:Lcomp}
\end{figure}

\begin{figure}
    \centering
    \includegraphics[width = 5 in]{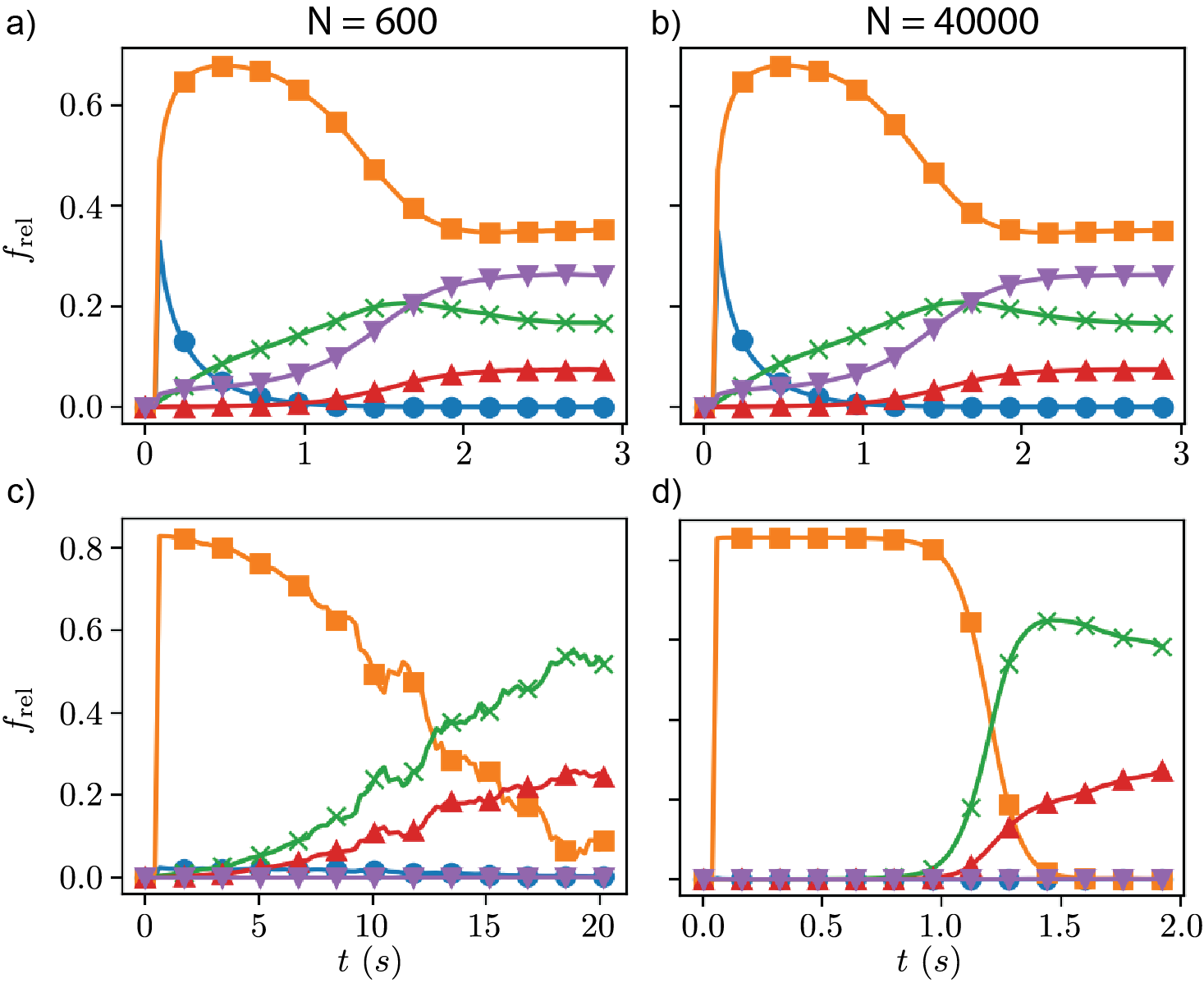}
    \caption{Relative reaction frequencies for the two different sets of rate constants for both large and small $N$. Orange ($\blacksquare$) corresponds to monomer addition, blue ($\bullet$) to primary nucleation, green ($\textbf{x}$) to fragmentation, red ($\blacktriangle$) to coagulation and purple ($\blacktriangledown$) to monomer subtraction. For \textbf{set 2} (a and b), there is no change in the reactions even for very small $N$. For \textbf{set 1} (c and d), not only do more reactions become relevant during the growth phase at small $N$, there are significant fluctuations in the reaction rates.}
    \label{fig:2setReacs}
\end{figure}

System volume has a significant effect on fluctuations of the moments of the distribution (polymer number and polymer mass), but also on the overall rate of mass production as well as the evolution of the average length of polymers. As mentioned earlier, for certain choices of rate constants the half-time of the reaction may increase or stay roughly the same as volume is decreased. In all cases, fluctuations increase with decreasing volume but the magnitude of fluctuations depends strongly on the rate constants themselves. Fig. \ref{fig:mass_dev} shows plots of the average polymerized mass as a function of time for \textbf{set 2}, normalized by the total mass of the system. For large values of $N$, i.e. larger volumes, deviations from the average are small, growing larger as $N$ is decreased. As expected, the value of the relative deviation is proportional to $1/\sqrt{N}$. The fluctuations in mass peak when the rate of change of mass is at its maximum, which corresponds with the time period in which the most reactions are occurring. Later, they reach an equilibrium value in which the total number of reactions occurring is smaller and roughly stable. It should be noted that for this set the reaction proceeds at the same time scale for even very small values of $N$.
Fig \ref{fig:bigfluc} shows the same plots for \textbf{set 1}. In this case, fluctuations are much larger, and can in fact be larger than the average mass. The time-scale of the reaction increases significantly as $N$ becomes smaller, corresponding to the half-time of the reaction increasing with $1/N$.
Another interesting result is presented in Fig. \ref{fig:Lcomp}. For \textbf{set 2}, $L(t)$ peaks very low (roughly $5$) and is almost identical for every value of $N$. For \textbf{set 1}, however, not only does the peak value of $L$ decrease, which would be expected as for small enough values of $N$ it could not possibly reach the same value, but the actual shape of $L(t)$ changes. For large $N$, $L(t)$ peaks sharply early on in the growth phase before rapidly falling back to smaller values. As $N$ is decreased, this peak broadens and eventually flattens at small enough values of $N$. This would seem to imply that the reaction mechanisms have very different behavior at different values of $N$. 
Fig \ref{fig:2setReacs} illustrates this clearly. For \textbf{set 2} (Figs. \ref{fig:2setReacs}($a$) and ($b$)), the reactions proceed almost identically for both $N=40000$ and $N=600$. For \textbf{set 1} (Figs. \ref{fig:2setReacs}($c$) and ($d$)), however, more reactions are involved during the growth phase for $N=600$ than there are for $N=40000$, and the reaction rates fluctuate fairly significantly. This relative increase in importance of mechanisms other than monomer addition, particularly fragmentation, would explain the flattening of the peak in $L(t)$.

\subsection{Average versus Individual-run Behavior}

\begin{figure}[t]
    \centering
    \includegraphics[width = 3.5 in]{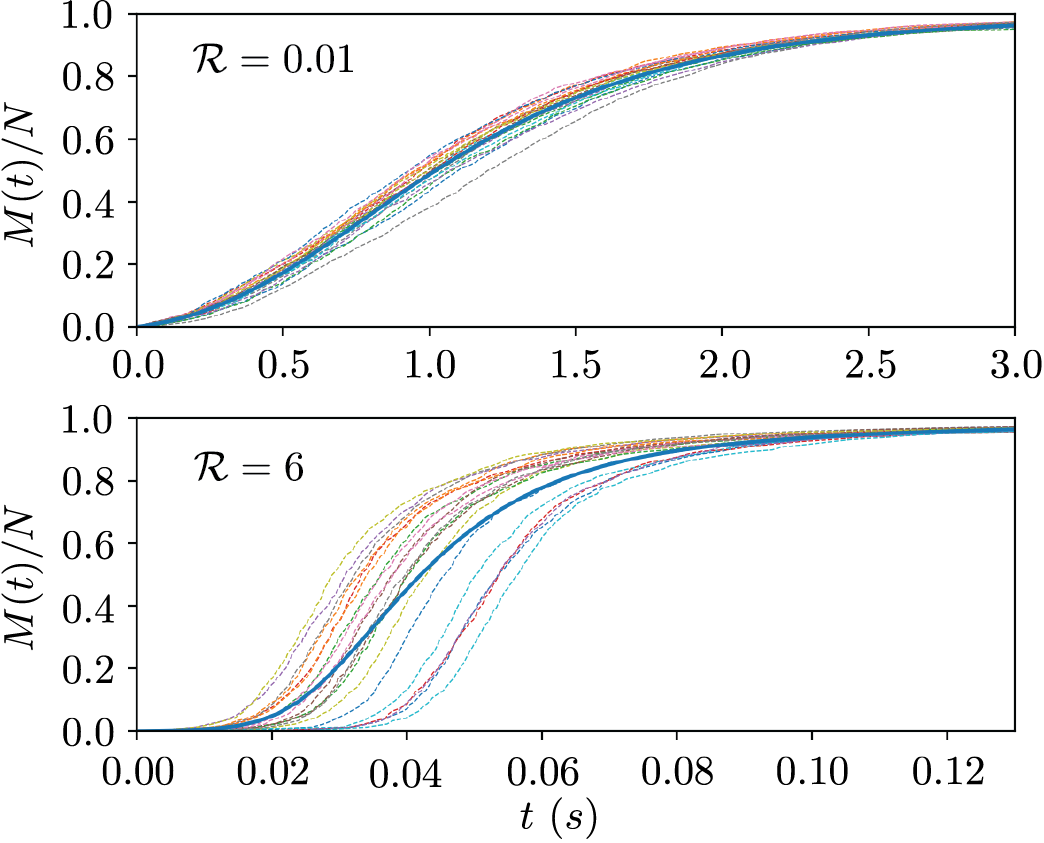}
    \caption{Local dynamics changing as $\mathcal{R}$ increases. The solid line represents the average while dashed lines represent individual stochastic simulation runs. Simulation were run with fixed parameters $N=2000$, $c_0 = \SI{10}{\micro\Molar}$, $a=\SI{2}{\micro\Molar^{-1} s^{-1}}$, $b=\SI{0.1}{s^{-1}}$, $k_a=k_b=\SI{0}{}$, $k_n = \SI{0.005}{\micro\Molar^{-1} s^{-1}}$ and $n_2=n_c=2$.}
    \label{fig:R_ind}
\end{figure}

In addition to giving access to fluctuations, stochastic simulations allow for direct comparison of individual reaction pathways to the average behavior of a set of simulations. This is analogous to a comparison of bulk behavior to single-molecule behavior in the field of single-molecule experiments, the latter of which is much more difficult to access experimentally. For this section, we ran a sweep of the ratio of parameters 

\begin{equation}
\mathcal{R}=c_0^{n_2-1}\frac{k_2}{a},
\end{equation}

where $c_0^{n_2-1}$ on the right hand side makes the ratio dimensionless, to gain insight on how individual reaction pathways may change as parameters are varied. Fig. \ref{fig:R_ind} shows two sets of simulations at different values of $\mathcal{R}$ by varying $k_2$ and keeping the rest of the  rate constants the same. It is clear that when secondary nucleation is relatively unimportant, the individual runs of the simulation behave much like their average. However, when this ratio becomes large, the individual behavior can deviate quite significantly from the average. 



Additionally, for low $\mathcal{R}$, it turns out that the distribution of halftime about the mean value is close to normal, while for larger values it becomes skewed.


\subsection{Crowders Change Dynamics}
\begin{figure}
    \centering
    \includegraphics[width = 5 in]{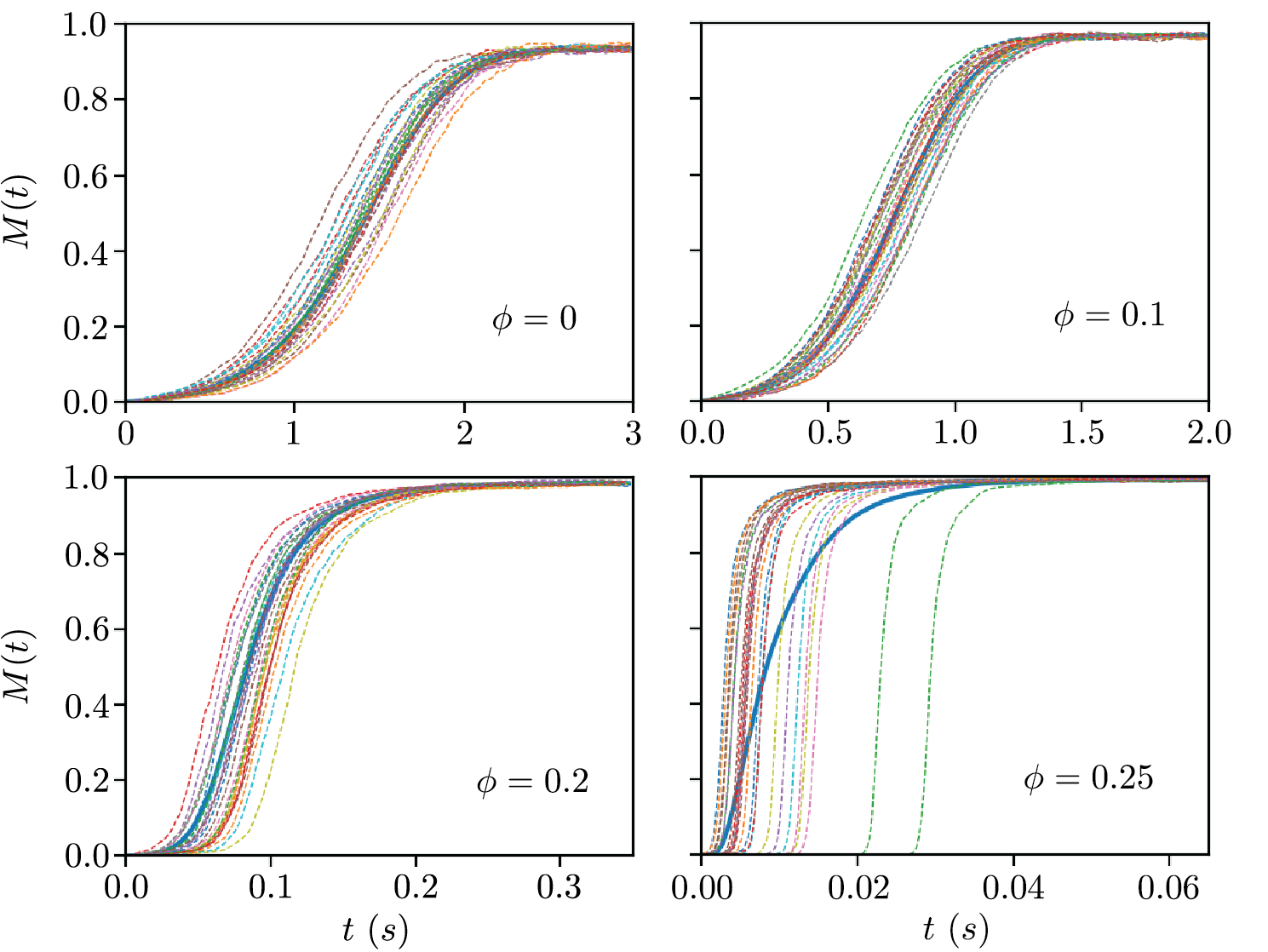}
    \caption{Local dynamics changing as crowder volume fraction is increased. Individual stochastic runs are represented by dashed lines, the average by a solid line. For no crowders and low crowder concentrations, the individual runs behave similarly to the average. At large $\phi$, however, the individual runs behave very differently from the average.}
    \label{fig:local_crowds}
\end{figure}

\begin{figure}
    \centering
    \includegraphics[width = 5 in]{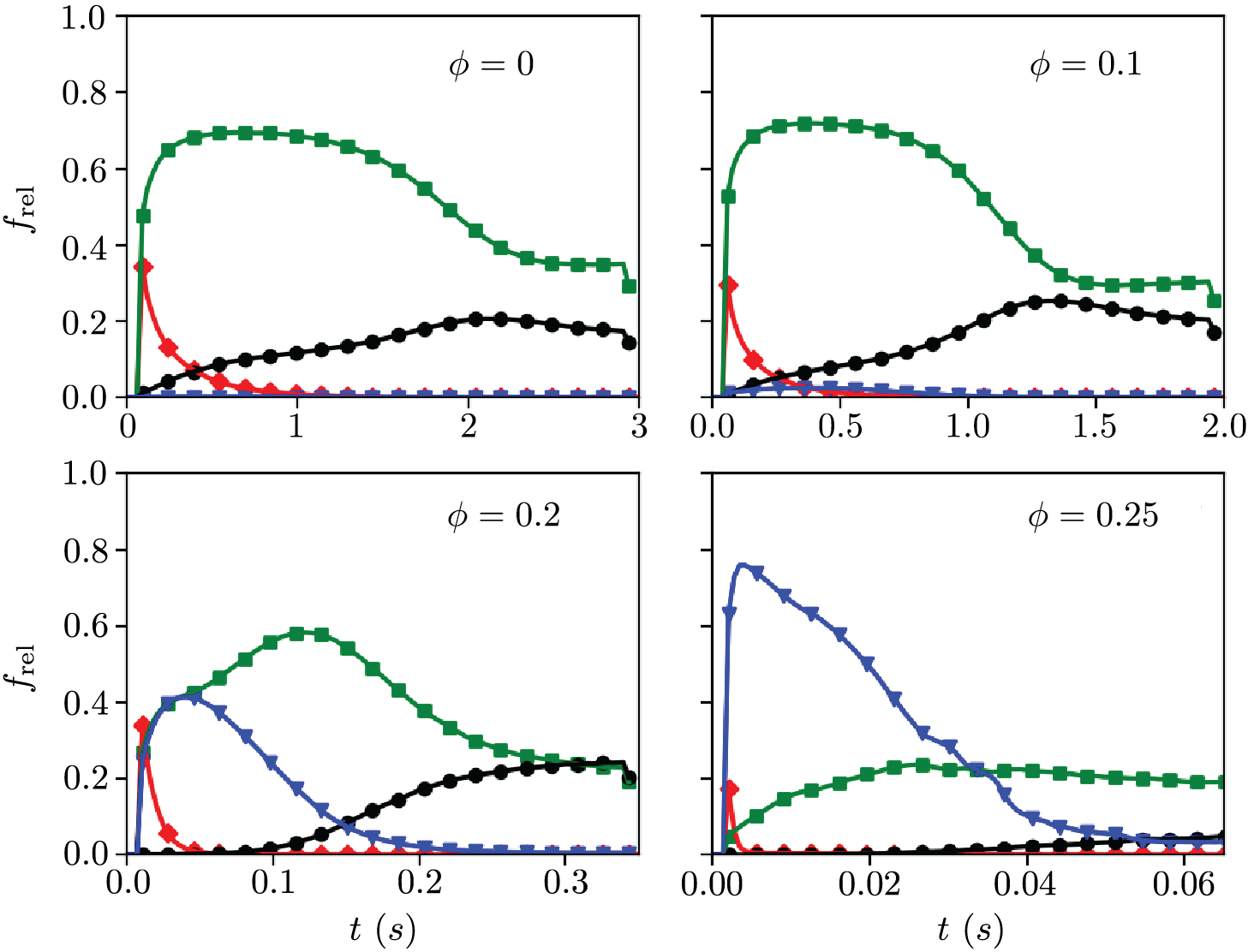}
    \caption{Relative frequency of the various mechanisms of growth. Curves shown are secondary nucleation ($\blacktriangledown$), primary nucleation ($\sqdiamond$), fragmentation ($\bullet$) and monomer addition ($\blacksquare$). As $\phi$ increases, secondary nucleation begins to supress all other mechanisms. At intermediate values, it is clear that there is competition between multiple growth mechanisms.}
    \label{fig:local_reactions}
\end{figure}

A more physiologically relevant example of the present model is to see how the presence of crowder molecules can change the local behavior. Fig. \ref{fig:local_crowds} shows that as the volume fraction, $\phi$, of crowders increases, we see similar spread in local behavior compared to the average as we did when directly changing the rates. This is because the growth rates are directly affected by $\phi$ as seen in the theory section. 
From Fig. \ref{fig:local_reactions}, you can see that secondary nucleation completely dominates the growth process as $\phi$ is increased. However, before it can occur, an incubation period exists, as shown in the case of $\phi = 0.25$ of Fig. \ref{fig:local_reactions}. Hence, in the individual runs at large $\phi$, there is a short period of no growth before the first primary nucleation event occurs, followed by a rapid explosive period of growth once a polymer has formed and the auto-catalytic secondary-nucleation process can occur. These simulations were run with $n_c=n_2=2$, and the effect is generally more exaggerated when $n_2>n_c$. This is a purely stochastic phenomenon, as fluctuations in the first-passage time of primary nucleation (as well as monomer addition in the case of $n_2>n_c$) leads to the spread of the individual runs, producing an average that does not represent the individual reaction pathways of the system. Moreover, it is clear that the presence of crowders can magnify these differences by increasing certain reaction propensities (e.g secondary nucleation) more than others (such as merging or addition).

\begin{figure}
    \centering
    \includegraphics[width = 3.5 in]{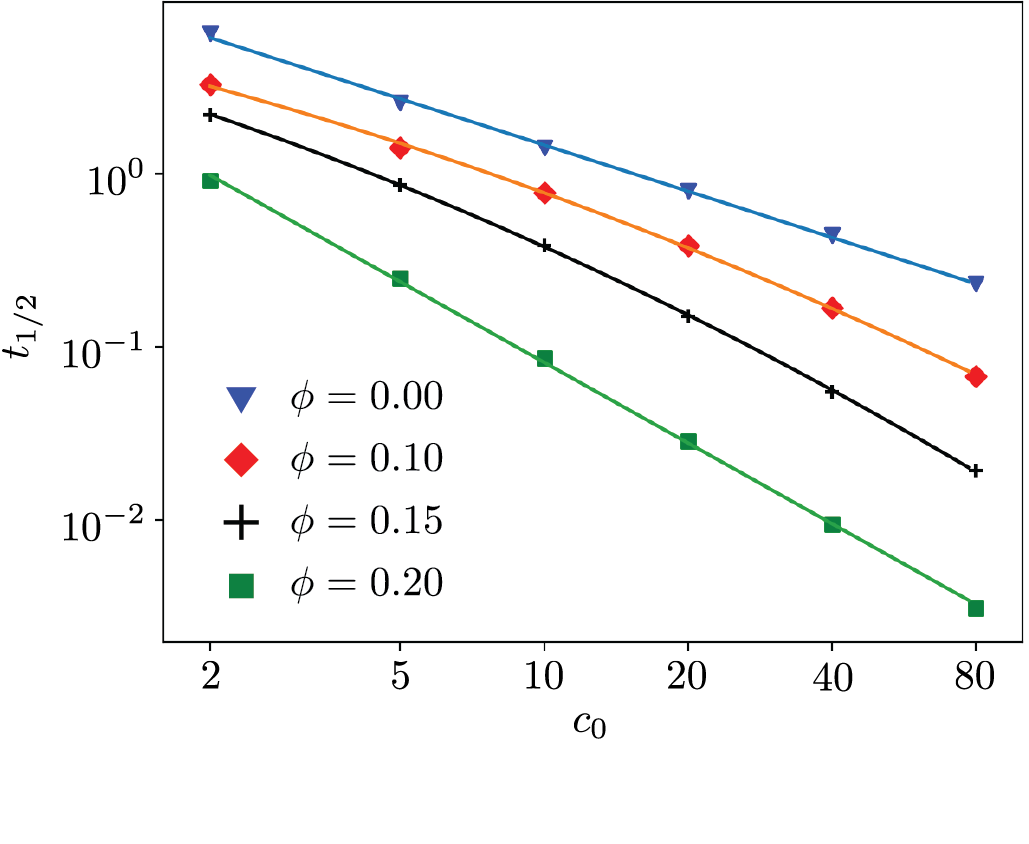}
    \caption{Scaling law comparison of the same system with different volume fraction ($\phi$) of crowders. For the crowder-less case, the slope is close to $1$ corresponding to nucleation-elongation. For $\phi=0.2$, it is close to $1.5$ corresponding to secondary nucleation. In between, there is competition between nucleation-elongation and secondary nucleation as the dominant mechanisms shift.}
    \label{fig:cr_scaling}
\end{figure}

In other words, beyond increasing the overall rate of the reaction, the actual way in which the polymers grow is changed significantly. This can be shown in terms of the scaling law analysis as well. Fig. \ref{fig:cr_scaling} shows how the scaling law governing the crowderless reaction is significantly different from that with even fairly low $\phi$. The scaling laws here agree with the reaction picture, in that for $\phi=0$ the slope is close to $1$ (corresponding to nucleation-elongation. $\gamma=-n_c/2$) and for $\phi=0.2$ the slope is $1.5$ (secondary nucleation dominating. $\gamma=-(n_2+1)/2$). One implication of this is that knowledge of the dynamics of an aggregating system \textit{in vitro} does not necessarily translate to the same protein aggregating \textit{in vivo}, where much of the system is occupied by molecules which do not participate in the reaction other than to exclude volume, resulting in the entropic effects.

\subsection{Comparison With the PM-model}

\begin{figure}
    \centering
    \includegraphics[width = 6.0 in]{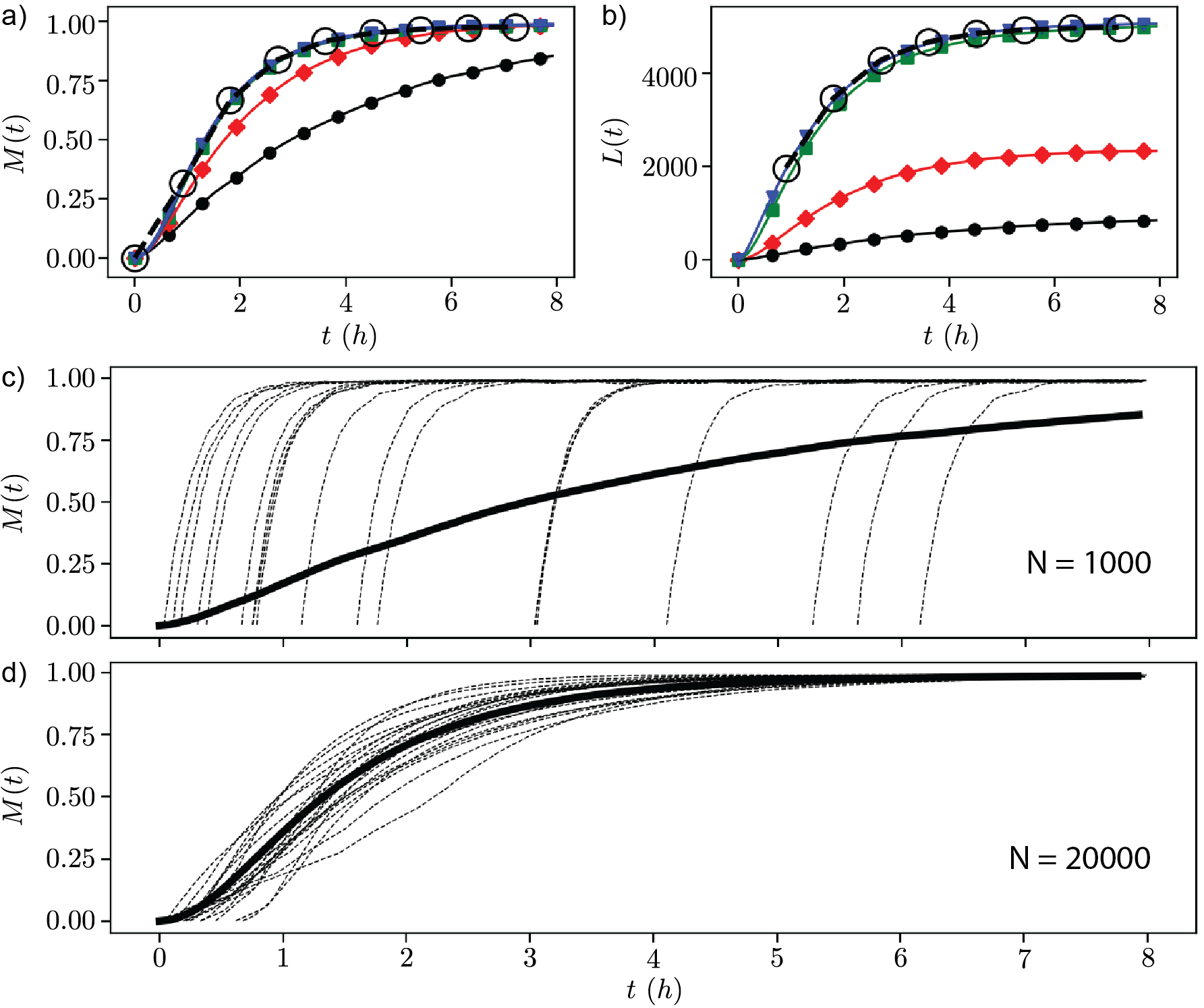}
    \caption{The fraction of mass contained in polymers (a) and the average length of polymers (b) compared with the PM-model (dashed). Stochastic simulations were done with N = 1000 ($\bullet$), 2500 ($\sqdiamond$), 10000 ($\blacksquare$), 20000 ($\blacktriangledown$). In (a) and (b), open circles refer to the mean value obtained using the PM model. In (c) and (d), comparison of the individual runs (dashed) with the average (solid) of the stochastic model. For large N, the individual runs closely resemble the average, which fits the PM-model prediction. For smaller N, however, not only do the results differ drastically from the continuous model, the individual behavior changes relative to the average. Simulations were run using the Oosawa model with parameters from Schreck et al.\cite{schreck20}, obtained by fitting ThT data for actin in the presence of dextran from Rosin et al.\cite{winterDex}, $a=\SI{911}{\micro\Molar^{-1}s^{-1}}$, $b=\SI{50}{s^{-1}}$, $k_n=\SI{1.02e-4}{\micro\Molar^{-2}s^{-1}}$ and $n_c=3$.}
    \label{fig:M60vsMP}
\end{figure}

In order to  show more clearly the effects of low particle number, we compared the stochastic approach to the moment-closure 
approximation of the reaction-rate-equation approach. The latter will be called the PM-model\cite{knowles-09} in the present article. This model reduces a large set of rate equations for the concentration of each species to three closed differential equations. One for the monomer concentration, $c_1(t)$, and one each for the first two moments of the distribution of aggregates: the number of polymers, $P(t)$, and the mass contained in polymers, $M(t)$. Additionally, the average length of polymers, $L(t)$, is computed as the ratio $M(t)/P(t)$. For a more detailed description, we refer the reader to the references.\cite{michaels14,hong17,schreck20,meisl14} Figs. \ref{fig:M60vsMP}(a) and (b) show comparisons of $M(t)$ and $L(t)$ for the PM-model (fitted to experimental data in Schreck et al.\cite{schreck20}) as well as stochastic simulation at different values of $N$. It is clear that for large $N$, the stochastic results agree with the bulk behavior predicted by the PM-model, but differ significantly as $N$ decreases. For lower values of $N$, the overall rate of production of mass and asymptotic length are much decreased. Of course, for certain small values of $N$, depending on chosen rate constants, the average length predicted by stochastic dynamics could not possibly agree with that of the PM-model. It is important to note, however, that $L$ converges concurrently with $M$ as $N$ is increased. An interpretation of this is that, for stochastic dynamics to agree with bulk methods, $N$ must be large enough for the length profiles to agree. 

Figs. \ref{fig:M60vsMP}(c) and (d) show even more interesting behavior. For smaller $N$, the individual stochastic runs actually reach their equilibrium mass values \textit{more rapidly}, following initial nucleation, than do the individual runs for large $N$. So the reduction in the rate of relative mass production is purely due to the stochastic nature of the system: nucleation events are more spread out and, on average, take longer to occur. Additionally, at larger $N$, more than one nucleation event generally occurs as evidenced by kinks in the individual runs. When considering that the number of molecules in the simulation is analogous to the volume of the system at fixed concentration, this implies that the dynamics may change significantly depending on system volume.




\section{Discussion and Conclusions}

In summary, we have introduced a powerful stochastic method based on the Gillespie SSA to study the time evolution of the relative frequencies of aggregation reaction mechanisms. We tested this method by comparing with the scaling law treatment of Meisl et al.\cite{meisl2016molecular} to confirm that the dominant mechanisms predicted by the scaling exponent, $\gamma$, agreed with the relative reaction frequencies of those mechanisms. In particular, we showed that as the scaling exponent increased, competition between mechanisms was seen as the relative frequency of monomer addition was overcome by secondary nucleation as $c_0$ increased. Additionally, detail as to which mechanisms become important and at what time during the reaction can give insight into why observable quantities, such as polymer mass, behave as they do. The method provides great detail into the behavior of reacting systems and, in principle, can be used for any stochastic system where knowledge of the importance of particular reactions or events as they change over time is desired.

We showed that the halftime of $M(t)$ is not always proportional to $1/N$ (or $1/V$), as shown by Tiwari and van der Schoot\cite{tiwari16}, and in fact can reach a thermodynamic limit where increasing $N$ has no effect on $t_{1/2}$. This effect, along with the behavior of $\sigma_{1/2}$, is strongly dependent on the values of the rate constants. Accordingly, for a choice of rate constants where $t_{1/2}$ does not increase with decreasing volume, the time-scale of $M(t)$ does not change as volume is decreased, whereas it can increase quite significantly for another set of rate constants. By looking at the average polymer length over time, we showed that simulations with rate constants which favored smaller polymers did not show changing results for the range of $N$ used in this study, while those using rate constants which favored very long polymers had dramatic changes in dynamics. This is further confirmed by the reaction frequencies for longer polymer-forming rate constants fluctuating more significantly, and more mechanisms being important in the reaction at smaller values of $N$. In other words, the dynamics can be very different for small volumes. Physically, this is consistent with smaller volumes being less conducive to very large polymers growing. This implies that, when fitting models to experimental data in order to predict behavior within living cells, one should also fit $L(t)$ to data on the average length of polymers, or at the very least bias the fits to achieve a best guess of the actual length profiles, as was done in Schreck et al.\cite{schreck20}. That atomic force microscope (AFM) measurements of polymer length can be used in addition to ThT measurements to obtain more robust estimates of rate constants was previously pointed out by Schreck and Yuan.\cite{Schreck13}

We also compared individual stochastic reaction pathways with the average value calculated from many runs for a range of parameter values, as well as in the presence of crowders. We showed that for certain parameters, the individual runs can look very different than the average. In the case studied, when secondary nucleation was relatively unimportant compared to monomer addition, the individual runs looked similar to the average. When secondary nucleation was made more important, however, the individual runs differed greatly from the average. This reflects both the change in reaction dynamics and the skewness in the distribution of $t_{1/2}$. When crowders were included, this change was even more dramatic, and plots of the reaction frequencies indeed confirm that secondary nucleation could become completely dominant. 

Furthermore, we compared the stochastic approach to the continuous, PM-model. For large values of $N$ the two models are in good agreement, but diverge significantly as $N$ decreases. Again, we saw that $L(t)$ not being in agreement for the two models was indicative of this divergence, further confirming the importance of having experimental data on the lengths of polymers. Additionally, analysis of the individual runs shows that, for small values of $N$, the local behavior is vastly different than the average behavior. This means that the local behavior effect mentioned previously can be caused not only by the presence of crowders, but by decreasing the reaction volume. The specific model compared was the Oosawa model without secondary nucleation, so the effect is present even in the simplest of models provided they have more than one mechanism of growth.

Lastly, to our stochastic kinetic simulator we can add other reaction mechanisms to different degrees of sophistication, depending on the system that we are investigating. An important next step is to apply our methods to protein aggregation systems that have been studied experimentally, especially, \emph{in vivo}. For such systems, we should first fit the observed $M(t)$ and/or $L(t)$ curves by varying rate constants and other parameters \cite{Schreck13,schreck20}. With the set of constants determined, the present stochastic scheme can be used to provide valuable information on the fluctuations, the reaction dynamics, and pathways involved in the system and how they vary with the changes of system volume and the amount of crowders.

All stochastic simulations were performed using \textit{popsim}\cite{popsim}, a browser-based program developed by the authors for this study. It is available for use at www.popsim.xyz. The source code is available at https://github.com/jljorgenson18/popsim.

\section{Acknowledgements}

We thank Prof. Frank Ferrone for stimulating discussion, Karsten Chu for useful discussion at the early stage of the work. It is late, but we still want to dedicate this work to Prof. William P. Reinhardt for his influence on our research in general. The authors have no funding institutions/support to report.

\bibliography{stochastic_kinetics}

\providecommand{\latin}[1]{#1}
\makeatletter
\providecommand{\doi}
  {\begingroup\let\do\@makeother\dospecials
  \catcode`\{=1 \catcode`\}=2 \doi@aux}
\providecommand{\doi@aux}[1]{\endgroup\texttt{#1}}
\makeatother
\providecommand*\mcitethebibliography{\thebibliography}
\csname @ifundefined\endcsname{endmcitethebibliography}
  {\let\endmcitethebibliography\endthebibliography}{}
\begin{mcitethebibliography}{49}
\providecommand*\natexlab[1]{#1}
\providecommand*\mciteSetBstSublistMode[1]{}
\providecommand*\mciteSetBstMaxWidthForm[2]{}
\providecommand*\mciteBstWouldAddEndPuncttrue
  {\def\EndOfBibitem{\unskip.}}
\providecommand*\mciteBstWouldAddEndPunctfalse
  {\let\EndOfBibitem\relax}
\providecommand*\mciteSetBstMidEndSepPunct[3]{}
\providecommand*\mciteSetBstSublistLabelBeginEnd[3]{}
\providecommand*\EndOfBibitem{}
\mciteSetBstSublistMode{f}
\mciteSetBstMaxWidthForm{subitem}{(\alph{mcitesubitemcount})}
\mciteSetBstSublistLabelBeginEnd
  {\mcitemaxwidthsubitemform\space}
  {\relax}
  {\relax}

\bibitem[{L. Blanchoin, R. Boujemaa-Paterski, C. Sykes, J.
  Plastino.}(2014)]{actin}
{L. Blanchoin, R. Boujemaa-Paterski, C. Sykes, J. Plastino.}, Actin Dynamics,
  Architecture, and Mechanics in Cell Motility. \emph{Physiol. Rev}
  \textbf{2014}, \emph{94}, 235--63\relax
\mciteBstWouldAddEndPuncttrue
\mciteSetBstMidEndSepPunct{\mcitedefaultmidpunct}
{\mcitedefaultendpunct}{\mcitedefaultseppunct}\relax
\EndOfBibitem
\bibitem[Conde and Caceres(2009)Conde, and Caceres]{microtubule}
Conde,~C.; Caceres,~A. Microtubule Assembly, Organization and Dynamics in Axons
  and Dendrites. \emph{Nat. Rev. Neurosci.} \textbf{2009}, \emph{10},
  319--32\relax
\mciteBstWouldAddEndPuncttrue
\mciteSetBstMidEndSepPunct{\mcitedefaultmidpunct}
{\mcitedefaultendpunct}{\mcitedefaultseppunct}\relax
\EndOfBibitem
\bibitem[{T.P.J. Knowles, M. Vendruscolo, C. M.
  Dobson}(2015)]{Knowles-PhysToday}
{T.P.J. Knowles, M. Vendruscolo, C. M. Dobson}, The Physical Basis of Protein
  Misfolding Disorders. \emph{Phys. Today} \textbf{2015}, \emph{68}, 36\relax
\mciteBstWouldAddEndPuncttrue
\mciteSetBstMidEndSepPunct{\mcitedefaultmidpunct}
{\mcitedefaultendpunct}{\mcitedefaultseppunct}\relax
\EndOfBibitem
\bibitem[Yuan and Zhou(2017)Yuan, and Zhou]{yuan17}
Yuan,~J.~M., Zhou,~H.~X., Eds. \emph{Biophysics and Biochemistry of Protein
  Aggregation: Experimental and Theoretical Studies on Folding, Misfolding, and
  Self-Assembly of Amyloidogenic Peptides}; World Scientific, 2017\relax
\mciteBstWouldAddEndPuncttrue
\mciteSetBstMidEndSepPunct{\mcitedefaultmidpunct}
{\mcitedefaultendpunct}{\mcitedefaultseppunct}\relax
\EndOfBibitem
\bibitem[{T. Knowles, C. Waudby, G. Devlin, S. Cohen, A. Agguzzi, M.
  Vendruscolo, E. Terentjev, M. Welland, C. Dobson}(2009)]{knowles-09}
{T. Knowles, C. Waudby, G. Devlin, S. Cohen, A. Agguzzi, M. Vendruscolo, E.
  Terentjev, M. Welland, C. Dobson}, An Analytical Solution to the Kinetics of
  Breakable Filament Assembly. \emph{Science} \textbf{2009}, \emph{326},
  1533--1537\relax
\mciteBstWouldAddEndPuncttrue
\mciteSetBstMidEndSepPunct{\mcitedefaultmidpunct}
{\mcitedefaultendpunct}{\mcitedefaultseppunct}\relax
\EndOfBibitem
\bibitem[{T.C.T Michaels and T.P.J. Knowles}(2014)]{michaels14}
{T.C.T Michaels and T.P.J. Knowles}, Role of Filament Annealing in the Kinetics
  and Thermodynamics of Nucleated Polymerization. \emph{J. Chem. Phys.}
  \textbf{2014}, \emph{140}, 214904\relax
\mciteBstWouldAddEndPuncttrue
\mciteSetBstMidEndSepPunct{\mcitedefaultmidpunct}
{\mcitedefaultendpunct}{\mcitedefaultseppunct}\relax
\EndOfBibitem
\bibitem[{L. Hong, C. F. Lee, Y. J. Huang}(2017)]{hong17}
{L. Hong, C. F. Lee, Y. J. Huang}, In \emph{Biophysics and Biochemistry of
  Protein Aggregation: Experimental and Theoretical Studies on Folding,
  Misfolding, and Self-Assembly of Amyloidogenic Peptides}; Yuan,~J.~M.,
  Zhou,~H.~X., Eds.; World Scientific, 2017; pp 113--186\relax
\mciteBstWouldAddEndPuncttrue
\mciteSetBstMidEndSepPunct{\mcitedefaultmidpunct}
{\mcitedefaultendpunct}{\mcitedefaultseppunct}\relax
\EndOfBibitem
\bibitem[Urbanc(2017)]{urbanc17}
Urbanc,~B. In \emph{Biophysics and Biochemistry of Protein Aggregation:
  Experimental and Theoretical Studies on Folding, Misfolding, and
  Self-Assembly of Amyloidogenic Peptides}; Yuan,~J.~M., Zhou,~H.~X., Eds.;
  World Scientific, 2017; pp 1--50\relax
\mciteBstWouldAddEndPuncttrue
\mciteSetBstMidEndSepPunct{\mcitedefaultmidpunct}
{\mcitedefaultendpunct}{\mcitedefaultseppunct}\relax
\EndOfBibitem
\bibitem[{C. Frieden}(2007)]{frieden07}
{C. Frieden}, Protein Aggregation Processes: In Search of the Mechanism.
  \emph{Protein Sci.} \textbf{2007}, \emph{16}, 2334--2344\relax
\mciteBstWouldAddEndPuncttrue
\mciteSetBstMidEndSepPunct{\mcitedefaultmidpunct}
{\mcitedefaultendpunct}{\mcitedefaultseppunct}\relax
\EndOfBibitem
\bibitem[{R.M. Murphy}(2007)]{murphy07}
{R.M. Murphy}, Kinetics of Amyloid Formation and Membrane Interaction with
  Amyloidogenic Proteins. \emph{Biochem. Biophys. Acta} \textbf{2007},
  \emph{1768}, 1923--1934\relax
\mciteBstWouldAddEndPuncttrue
\mciteSetBstMidEndSepPunct{\mcitedefaultmidpunct}
{\mcitedefaultendpunct}{\mcitedefaultseppunct}\relax
\EndOfBibitem
\bibitem[{J. Gillam and C. MacPhee}(2013)]{gillam13}
{J. Gillam and C. MacPhee}, Modelling Amyloid Fibril Formation Kinetics:
  Mechanisms of Nucleation and Growth. \emph{J. Phys. Condens. Matter.}
  \textbf{2013}, \emph{25}, 373101\relax
\mciteBstWouldAddEndPuncttrue
\mciteSetBstMidEndSepPunct{\mcitedefaultmidpunct}
{\mcitedefaultendpunct}{\mcitedefaultseppunct}\relax
\EndOfBibitem
\bibitem[Schreck and Yuan(2013)Schreck, and Yuan]{Schreck13}
Schreck,~J.~S.; Yuan,~J.~M. A Kinetic Study of Amyloid Formation: Fibril Growth
  and Length Distributions. \emph{J. Phys. Chem. B.} \textbf{2013}, \emph{117},
  6574--83\relax
\mciteBstWouldAddEndPuncttrue
\mciteSetBstMidEndSepPunct{\mcitedefaultmidpunct}
{\mcitedefaultendpunct}{\mcitedefaultseppunct}\relax
\EndOfBibitem
\bibitem[{F. A. Ferrone, J. Hofrichter, W. A. Eaton}(1980)]{ferrone80}
{F. A. Ferrone, J. Hofrichter, W. A. Eaton}, Kinetic Studies on
  Photolysis-induced Gelation of Sickle Cell Hemoglobin Suggest A New
  Mechanism. \emph{Biophys. J.} \textbf{1980}, \emph{32}, 361\relax
\mciteBstWouldAddEndPuncttrue
\mciteSetBstMidEndSepPunct{\mcitedefaultmidpunct}
{\mcitedefaultendpunct}{\mcitedefaultseppunct}\relax
\EndOfBibitem
\bibitem[Hofrichter(1986)]{hofrichter}
Hofrichter,~J. Kinetics of Sickle Hemoglobin Polymerization. III. Nucleation
  Rates Determined From Stochastic Fluctuations in Polymerization Progress
  Curves. \emph{J. Mol. Biol.} \textbf{1986}, \emph{189}\relax
\mciteBstWouldAddEndPuncttrue
\mciteSetBstMidEndSepPunct{\mcitedefaultmidpunct}
{\mcitedefaultendpunct}{\mcitedefaultseppunct}\relax
\EndOfBibitem
\bibitem[{T. P. J. Knowles, D. A. White, A. R. Abate, J. J. Agresti, S. I. A.
  Cohen, R. A. Sperling, E. J. De Genst, C. M. Dobson, D. A.
  Weitz}(2011)]{knowles11}
{T. P. J. Knowles, D. A. White, A. R. Abate, J. J. Agresti, S. I. A. Cohen, R.
  A. Sperling, E. J. De Genst, C. M. Dobson, D. A. Weitz}, Observation of
  Spatial Propagation of Amyloid Assembly From Single Nuclei. \emph{Proc. Natl.
  Acad. Sci. U.S.A.} \textbf{2011}, \emph{108}, 14746--14751\relax
\mciteBstWouldAddEndPuncttrue
\mciteSetBstMidEndSepPunct{\mcitedefaultmidpunct}
{\mcitedefaultendpunct}{\mcitedefaultseppunct}\relax
\EndOfBibitem
\bibitem[{D. W. Colby, J. P. Cassady, G. C. Lin, V. M. Ingram, K. D.
  Wittrup}(2006)]{wittrup06}
{D. W. Colby, J. P. Cassady, G. C. Lin, V. M. Ingram, K. D. Wittrup},
  Stochastic Kinetics of Intracellular Huntingtin Aggregate Formation.
  \emph{Nat. Chem. Biol.} \textbf{2006}, \emph{2}, 319--23\relax
\mciteBstWouldAddEndPuncttrue
\mciteSetBstMidEndSepPunct{\mcitedefaultmidpunct}
{\mcitedefaultendpunct}{\mcitedefaultseppunct}\relax
\EndOfBibitem
\bibitem[Michaels \latin{et~al.}(2016)Michaels, Dear, Kirkegaard, Saar, Weitz,
  and Knowles]{michaels16}
Michaels,~T.~C.; Dear,~A.~J.; Kirkegaard,~J.~B.; Saar,~K.~L.; Weitz,~D.~A.;
  Knowles,~T.~P. Fluctuations in the Kinetics of Linear Protein Self-Assembly.
  \emph{Phys. Rev. Lett.} \textbf{2016}, \emph{116}, 258103\relax
\mciteBstWouldAddEndPuncttrue
\mciteSetBstMidEndSepPunct{\mcitedefaultmidpunct}
{\mcitedefaultendpunct}{\mcitedefaultseppunct}\relax
\EndOfBibitem
\bibitem[{J. Szavits-Nossan, K. Eden, R. J. Morris, C. E. MacPhee, M. R. Evans,
  R. J. Allen}(2014)]{Szavits14}
{J. Szavits-Nossan, K. Eden, R. J. Morris, C. E. MacPhee, M. R. Evans, R. J.
  Allen}, Inherent Variability in the Kinetics of Autocatalytic Protein
  Self-assembly. \emph{Phys. Rev. Lett.} \textbf{2014}, \emph{113},
  098101--1--5\relax
\mciteBstWouldAddEndPuncttrue
\mciteSetBstMidEndSepPunct{\mcitedefaultmidpunct}
{\mcitedefaultendpunct}{\mcitedefaultseppunct}\relax
\EndOfBibitem
\bibitem[Tiwari and van~der Schoot(2016)Tiwari, and van~der Schoot]{tiwari16}
Tiwari,~N.~S.; van~der Schoot,~P. Stochastic Lag Time in Nucleated Linear
  Self-Assembly. \emph{J. Chem. Phys.} \textbf{2016}, \emph{23},
  235101--1--13\relax
\mciteBstWouldAddEndPuncttrue
\mciteSetBstMidEndSepPunct{\mcitedefaultmidpunct}
{\mcitedefaultendpunct}{\mcitedefaultseppunct}\relax
\EndOfBibitem
\bibitem[Gillespie(1976)]{gillespie76}
Gillespie,~D.~T. A General Method For Numerically Simulating the Stochastic
  Time Evolution of Coupled Chemical Reactions. \emph{J. Comput. Phys.}
  \textbf{1976}, \emph{22}, 403--34\relax
\mciteBstWouldAddEndPuncttrue
\mciteSetBstMidEndSepPunct{\mcitedefaultmidpunct}
{\mcitedefaultendpunct}{\mcitedefaultseppunct}\relax
\EndOfBibitem
\bibitem[Gillespie(1977)]{gillespie77}
Gillespie,~D.~T. Exact Stochastic Simulation of Coupled Chemical Reactions.
  \emph{J. Phys. Chem.} \textbf{1977}, \emph{81}, 2340–61\relax
\mciteBstWouldAddEndPuncttrue
\mciteSetBstMidEndSepPunct{\mcitedefaultmidpunct}
{\mcitedefaultendpunct}{\mcitedefaultseppunct}\relax
\EndOfBibitem
\bibitem[Gillespie(1992)]{gillespie92}
Gillespie,~D.~T. A Rigorous Derivation of the Chemical Master Equation.
  \emph{Physica A} \textbf{1992}, \emph{188}, 404--25\relax
\mciteBstWouldAddEndPuncttrue
\mciteSetBstMidEndSepPunct{\mcitedefaultmidpunct}
{\mcitedefaultendpunct}{\mcitedefaultseppunct}\relax
\EndOfBibitem
\bibitem[Gillespie(2007)]{gillespie07}
Gillespie,~D.~T. Stochastic Simulation of Chemical Kinetics. \emph{Annu. Rev.
  Phys. Chem.} \textbf{2007}, \emph{58}, 35–55\relax
\mciteBstWouldAddEndPuncttrue
\mciteSetBstMidEndSepPunct{\mcitedefaultmidpunct}
{\mcitedefaultendpunct}{\mcitedefaultseppunct}\relax
\EndOfBibitem
\bibitem[Gibson and Bruck(2000)Gibson, and Bruck]{gibson00}
Gibson,~M.~A.; Bruck,~J. Efficient Exact Stochastic Simulation of Chemical
  Systems with Many Species and Many Channels. \emph{J. Phys. Soc. Japan}
  \textbf{2000}, \emph{40}, 1232–39\relax
\mciteBstWouldAddEndPuncttrue
\mciteSetBstMidEndSepPunct{\mcitedefaultmidpunct}
{\mcitedefaultendpunct}{\mcitedefaultseppunct}\relax
\EndOfBibitem
\bibitem[McQuarrie(1967)]{mcquarrie67}
McQuarrie,~D. Stochastic Approach to Chemical Kinetics. \emph{J. Appl. Probab.}
  \textbf{1967}, \emph{4}, 413–78\relax
\mciteBstWouldAddEndPuncttrue
\mciteSetBstMidEndSepPunct{\mcitedefaultmidpunct}
{\mcitedefaultendpunct}{\mcitedefaultseppunct}\relax
\EndOfBibitem
\bibitem[{T. C. T. Michaels, A. J. Dear, T. P. J. Knowles}(2018)]{michaels18}
{T. C. T. Michaels, A. J. Dear, T. P. J. Knowles}, Stochastic Calculus of
  Protein Filament Formation Under Spatial Confinement. \emph{New J. Phys.}
  \textbf{2018}, \emph{20}, 055007\relax
\mciteBstWouldAddEndPuncttrue
\mciteSetBstMidEndSepPunct{\mcitedefaultmidpunct}
{\mcitedefaultendpunct}{\mcitedefaultseppunct}\relax
\EndOfBibitem
\bibitem[Bridstrup and Yuan(2016)Bridstrup, and Yuan]{bridstrup16}
Bridstrup,~J.; Yuan,~J.~M. Effects of Crowders on the Equilibrium and Kinetic
  Properties of Protein Aggregation. \emph{Chem. Phys. Lett.} \textbf{2016},
  \emph{659}, 252--257\relax
\mciteBstWouldAddEndPuncttrue
\mciteSetBstMidEndSepPunct{\mcitedefaultmidpunct}
{\mcitedefaultendpunct}{\mcitedefaultseppunct}\relax
\EndOfBibitem
\bibitem[{D. Hall and A.P. Minton}(2002)]{hall02}
{D. Hall and A.P. Minton}, Effects of Inert Volume-excluding Macromolecules on
  Protein Fiber Formation. {I}. {E}quilibrium Models. \emph{Biophys. Chem.}
  \textbf{2002}, \emph{98}, 93--104\relax
\mciteBstWouldAddEndPuncttrue
\mciteSetBstMidEndSepPunct{\mcitedefaultmidpunct}
{\mcitedefaultendpunct}{\mcitedefaultseppunct}\relax
\EndOfBibitem
\bibitem[{D. Hall and A.P. Minton}(2004)]{hall04}
{D. Hall and A.P. Minton}, Effects of Inert Volume-excluding Macromolecules on
  Protein Fiber Formation. {II}. {K}inetic Models for Nucleated Fiber Growth.
  \emph{Biophys. Chem.} \textbf{2004}, \emph{107}, 299--316\relax
\mciteBstWouldAddEndPuncttrue
\mciteSetBstMidEndSepPunct{\mcitedefaultmidpunct}
{\mcitedefaultendpunct}{\mcitedefaultseppunct}\relax
\EndOfBibitem
\bibitem[{H. X. Zhou, G. Rivas, and A. P. Minton}(2008)]{zhou08}
{H. X. Zhou, G. Rivas, and A. P. Minton}, Macromolecular Crowding and
  Confinement: Biochemical, Biophysical, and Potential Physiological
  Consequences. \emph{Annu. Rev. Biophys.} \textbf{2008}, \emph{37},
  375--397\relax
\mciteBstWouldAddEndPuncttrue
\mciteSetBstMidEndSepPunct{\mcitedefaultmidpunct}
{\mcitedefaultendpunct}{\mcitedefaultseppunct}\relax
\EndOfBibitem
\bibitem[{J. S. Schreck, J. Bridstrup, and J. M. Yuan}(2017)]{schreck17}
{J. S. Schreck, J. Bridstrup, and J. M. Yuan}, In \emph{Biophysics and
  Biochemistry of Protein Aggregation: Experimental and Theoretical Studies on
  Folding, Misfolding, and Self-Assembly of Amyloidogenic Peptides};
  Yuan,~J.~M., Zhou,~H.~X., Eds.; World Scientific, 2017; pp 187--220\relax
\mciteBstWouldAddEndPuncttrue
\mciteSetBstMidEndSepPunct{\mcitedefaultmidpunct}
{\mcitedefaultendpunct}{\mcitedefaultseppunct}\relax
\EndOfBibitem
\bibitem[Schreck \latin{et~al.}(2020)Schreck, Bridstrup, and Yuan]{schreck20}
Schreck,~J.~S.; Bridstrup,~J.; Yuan,~J.-M. Investigating the Effects of
  Molecular Crowding on the Kinetics of Protein Aggregation. \emph{J. Phys.
  Chem. B} \textbf{2020}, \emph{124}, 9829--9839\relax
\mciteBstWouldAddEndPuncttrue
\mciteSetBstMidEndSepPunct{\mcitedefaultmidpunct}
{\mcitedefaultendpunct}{\mcitedefaultseppunct}\relax
\EndOfBibitem
\bibitem[Hong \latin{et~al.}(2020)Hong, Schreck, and
  {\v{S}}ulc]{hong2020understanding}
Hong,~F.; Schreck,~J.~S.; {\v{S}}ulc,~P. Understanding DNA interactions in
  Crowded Environments with a Coarse-Grained Model. \emph{Nucleic Acids Res}
  \textbf{2020}, \emph{48}, 10726--10738\relax
\mciteBstWouldAddEndPuncttrue
\mciteSetBstMidEndSepPunct{\mcitedefaultmidpunct}
{\mcitedefaultendpunct}{\mcitedefaultseppunct}\relax
\EndOfBibitem
\bibitem[{G. Meisl, J. B. Kirkegaard, P. Arosio, T. C. T. Michaels, M.
  Vendruscolo, C. M. Dobson, S. Linse, T. P. J. Knowles}(2016)]{meisl16-prot}
{G. Meisl, J. B. Kirkegaard, P. Arosio, T. C. T. Michaels, M. Vendruscolo, C.
  M. Dobson, S. Linse, T. P. J. Knowles}, Molecular Mechanisms of Protein
  Aggregation From Global Fitting of Kinetic Models. \emph{Nat. Protoc}
  \textbf{2016}, \emph{11}, 252--272\relax
\mciteBstWouldAddEndPuncttrue
\mciteSetBstMidEndSepPunct{\mcitedefaultmidpunct}
{\mcitedefaultendpunct}{\mcitedefaultseppunct}\relax
\EndOfBibitem
\bibitem[{F. Oosawa}(1975)]{oosawa}
{F. Oosawa}, \emph{Thermodynamics of the Polymerization of Protein}; Academia
  Press, 1975\relax
\mciteBstWouldAddEndPuncttrue
\mciteSetBstMidEndSepPunct{\mcitedefaultmidpunct}
{\mcitedefaultendpunct}{\mcitedefaultseppunct}\relax
\EndOfBibitem
\bibitem[{J.A.D. Wattis}(2006)]{Wattis}
{J.A.D. Wattis}, An Introduction to Mathematical Models of
  Coagulation-fragmentation Processes: A Discrete Deterministic Mean-field
  Approach. \emph{Physica D} \textbf{2006}, \emph{222}, 1--20\relax
\mciteBstWouldAddEndPuncttrue
\mciteSetBstMidEndSepPunct{\mcitedefaultmidpunct}
{\mcitedefaultendpunct}{\mcitedefaultseppunct}\relax
\EndOfBibitem
\bibitem[{R. Becker and W. D\"oring}(1935)]{BD}
{R. Becker and W. D\"oring}, Kinetische Behandlung der Keimbildung in
  Ubersattigten Dampfern. \emph{Ann. Phys} \textbf{1935}, \emph{24},
  719--52\relax
\mciteBstWouldAddEndPuncttrue
\mciteSetBstMidEndSepPunct{\mcitedefaultmidpunct}
{\mcitedefaultendpunct}{\mcitedefaultseppunct}\relax
\EndOfBibitem
\bibitem[{F. A. Ferrone, M. Ivanova, R. Jasuja}(2002)]{Ferrone02}
{F. A. Ferrone, M. Ivanova, R. Jasuja}, Heterogeneous Nucleation and Crowding
  in Sickle Hemoglobin: An Analytic Approach. \emph{Biophys. J.} \textbf{2002},
  \emph{82}, 399--406\relax
\mciteBstWouldAddEndPuncttrue
\mciteSetBstMidEndSepPunct{\mcitedefaultmidpunct}
{\mcitedefaultendpunct}{\mcitedefaultseppunct}\relax
\EndOfBibitem
\bibitem[{F. A. Ferrone, J. Hofrichter, W. A. Eaton}(1985)]{Ferrone85}
{F. A. Ferrone, J. Hofrichter, W. A. Eaton}, Kinetics of Sickle Hemoglobin
  Polymerization II. A Double Nucleation Mechanism. \emph{J. Mol. Biol.}
  \textbf{1985}, \emph{183}, 611--631\relax
\mciteBstWouldAddEndPuncttrue
\mciteSetBstMidEndSepPunct{\mcitedefaultmidpunct}
{\mcitedefaultendpunct}{\mcitedefaultseppunct}\relax
\EndOfBibitem
\bibitem[{G. Meisl, X. Yang, E. Hellstrand, B. Frohm, J. B. Kirkegraard, S. I.
  A. Cohen, C. M. Dobson, S. Linse, T. P. J. Knowles}(2014)]{meisl14}
{G. Meisl, X. Yang, E. Hellstrand, B. Frohm, J. B. Kirkegraard, S. I. A. Cohen,
  C. M. Dobson, S. Linse, T. P. J. Knowles}, Differences in Nucleation Behavior
  Underlie the Contrasting Aggregation Kinetics of the {A}$\beta$40 and
  {A}$\beta$42 Peptides. \emph{Proc. Natl. Acad. Sci. U.S.A.} \textbf{2014},
  \emph{111}, 9384--9389\relax
\mciteBstWouldAddEndPuncttrue
\mciteSetBstMidEndSepPunct{\mcitedefaultmidpunct}
{\mcitedefaultendpunct}{\mcitedefaultseppunct}\relax
\EndOfBibitem
\bibitem[{J. M. A. Padgett and S. Ilie}(2016)]{padgett16}
{J. M. A. Padgett and S. Ilie}, An Adaptive Tau-leaping Method for Stochastic
  Simulations of Reaction-diffusion Systems. \emph{AIP Adv} \textbf{2016},
  \emph{6}, 035217\relax
\mciteBstWouldAddEndPuncttrue
\mciteSetBstMidEndSepPunct{\mcitedefaultmidpunct}
{\mcitedefaultendpunct}{\mcitedefaultseppunct}\relax
\EndOfBibitem
\bibitem[{H. Eyring}(1935)]{eyring}
{H. Eyring}, The Activated Complex in Chemical Reactions. \emph{J. Chem. Phys.}
  \textbf{1935}, \emph{3}, 107--15\relax
\mciteBstWouldAddEndPuncttrue
\mciteSetBstMidEndSepPunct{\mcitedefaultmidpunct}
{\mcitedefaultendpunct}{\mcitedefaultseppunct}\relax
\EndOfBibitem
\bibitem[{M.A. Cotter}(1974)]{cotter74}
{M.A. Cotter}, Hard-rod Fluid: Scaled Particle Theory Revisited. \emph{Phys.
  Rev. A} \textbf{1974}, \emph{10}, 625\relax
\mciteBstWouldAddEndPuncttrue
\mciteSetBstMidEndSepPunct{\mcitedefaultmidpunct}
{\mcitedefaultendpunct}{\mcitedefaultseppunct}\relax
\EndOfBibitem
\bibitem[Minton(2014)]{Minton-14}
Minton,~A. The Effect of Time-dependent Macromolecular Crowding on the Kinetics
  of Protein Aggregation: {A} Simple Model for the Onset of Age-related
  Neurodegenerative Disease. \emph{Front. Phys.} \textbf{2014}, \emph{12}\relax
\mciteBstWouldAddEndPuncttrue
\mciteSetBstMidEndSepPunct{\mcitedefaultmidpunct}
{\mcitedefaultendpunct}{\mcitedefaultseppunct}\relax
\EndOfBibitem
\bibitem[{G. Meisl, L. Rajaj, S. A. I. Cohen, M. Pfammatter, A. Saric, E.
  Hellstrand, A. K. Buell, A. Aguzzi, M. Vendruscolo, C. M. Dobson, S. Linse,
  T. P. J. Knowles}(2017)]{meisl17}
{G. Meisl, L. Rajaj, S. A. I. Cohen, M. Pfammatter, A. Saric, E. Hellstrand, A.
  K. Buell, A. Aguzzi, M. Vendruscolo, C. M. Dobson, S. Linse, T. P. J.
  Knowles}, Scaling Behaviour and Rate-determining Steps In Filamentous
  Self-assembly. \emph{Chem. Sci.} \textbf{2017}, \emph{8}, 7087--7097\relax
\mciteBstWouldAddEndPuncttrue
\mciteSetBstMidEndSepPunct{\mcitedefaultmidpunct}
{\mcitedefaultendpunct}{\mcitedefaultseppunct}\relax
\EndOfBibitem
\bibitem[{C. Rosin, P.H. Schummel, R. Winter}(2015)]{winterDex}
{C. Rosin, P.H. Schummel, R. Winter}, Cosolvent and Crowding Effects on the
  Polymerization Kinetics of Actin. \emph{Phys. Chem. Chem. Phys.}
  \textbf{2015}, \emph{17}, 8330--7\relax
\mciteBstWouldAddEndPuncttrue
\mciteSetBstMidEndSepPunct{\mcitedefaultmidpunct}
{\mcitedefaultendpunct}{\mcitedefaultseppunct}\relax
\EndOfBibitem
\bibitem[Meisl \latin{et~al.}(2016)Meisl, Kirkegaard, Arosio, Michaels,
  Vendruscolo, Dobson, Linse, and Knowles]{meisl2016molecular}
Meisl,~G.; Kirkegaard,~J.~B.; Arosio,~P.; Michaels,~T.~C.; Vendruscolo,~M.;
  Dobson,~C.~M.; Linse,~S.; Knowles,~T.~P. Molecular Mechanisms of Protein
  Aggregation from Global Fitting of Kinetic Models. \emph{Nat. Protoc}
  \textbf{2016}, \emph{11}, 252--272\relax
\mciteBstWouldAddEndPuncttrue
\mciteSetBstMidEndSepPunct{\mcitedefaultmidpunct}
{\mcitedefaultendpunct}{\mcitedefaultseppunct}\relax
\EndOfBibitem
\bibitem[{J. Bridstrup, J. Jorgenson, J. M. Yuan}(2020)]{popsim}
{J. Bridstrup, J. Jorgenson, J. M. Yuan}, popsim: Browser-based Stochastic
  Simulation of Protein Aggregation. 2020;
  \url{https://github.com/jljorgenson18/popsim}\relax
\mciteBstWouldAddEndPuncttrue
\mciteSetBstMidEndSepPunct{\mcitedefaultmidpunct}
{\mcitedefaultendpunct}{\mcitedefaultseppunct}\relax
\EndOfBibitem
\end{mcitethebibliography}

\end{document}